\documentclass[9pt]{amsart}

\usepackage{graphicx, color}

\usepackage{algpseudocode}

\usepackage{algorithm}

\usepackage{subfig}

\usepackage{mathtools}
\mathtoolsset{showonlyrefs=true}

\newcommand{\R}{\mathbb{R}}

\newcommand{\Parent}{\mbox{Parent}}
\newcommand{\Face}{\mbox{Face}}
\newcommand{\Edge}{\mbox{Edge}}
\newcommand{\Center}{\mbox{Center}}
\newcommand{\Depth}{\mbox{Depth}}
\newcommand{\Extent}{\mbox{Extent}}

\newcommand{\Project}{\mbox{Project}}
\newcommand{\Distance}{\mbox{Distance}}

\theoremstyle{definition}
\newtheorem{theorem}{\sf Theorem}[section]
\newtheorem{definition}[theorem]{\sf Definition}
\newtheorem{remark}[theorem]{\sf Remark}
\newtheorem{lemma}[theorem]{\sf Lemma}

\calclayout

\begin{document}

\title{Efficient Exact Enumeration of Single-Source Geodesics on a Non-Convex Polyhedron}

\author[K.~Tateiri]{Kazuma Tateiri}\address[K.~Tateiri]{Graduate School of Information Science and Technology, Hokkaido University,Sapporo 060-0810, Japan}\email{kazuma.tateiri16@gmail.com}

\date{16 December 2023}

\keywords{Geodesics; enumeration; locally-shortest paths; polyhedral geodesics; discrete geodesics}

\begin{abstract}
In this paper, we consider enumeration of geodesics on a polyhedron, where a geodesic means locally-shortest path between two points. Particularly, we consider the following preprocessing problem: given a point $s$ on a polyhedral surface and a positive real number $r$, to build a data structure that enables, for any point $t$ on the surface, to enumerate all geodesics from $s$ to $t$ whose length is less than $r$. First, we present a naive algorithm by removing the trimming process from the MMP algorithm (1987). Next, we present an improved algorithm which is practically more efficient on a non-convex polyhedron, in terms of preprocessing time and memory consumption. Moreover, we introduce a \textit{single-pair geodesic graph} to succinctly encode a result of geodesic query. Lastly, we compare these naive and improved algorithms by some computer experiments.
\end{abstract}

\maketitle

\section{Introduction}
The shortest path problem is a fundamental problem in the field of discrete algorithm, which is also practically important and has been extensively investigated.
% %%
% A geodesic is ...
% A geodesic is said to be locally-shotest when ...., while it is globally-shotest when ... .
% In this paper, we focus on locally-shotest gedesics (only).
% %%
Particularly, there is a variety of work for the single-source shortest path problem on a polyhedral surface~\cite{MMP,CH,ICH,SVG}.
% A geodesic is a locally shortest path between two points on a polyhedral surface.
The \textit{single-source shortest path problem} on a polyhedron is defined as follows: An input to the problem consists of a polyhedron $\mathcal P$ and a point $s$ on $\mathcal P$. The task is to find a shortest path from $s$ to $t$ for all vertices $t$ of $\mathcal P$.
Mitchell, Mount and Papadimitriou~\cite{MMP} first
% studied this problem, and
presented an efficient algorithm for the problem, which computes the shortest path between two points on a polyhedron.
%Particularly,
Following~\cite{MMP}, there has been a variety of work for the single-source shortest path problem on a polyhedron, including
%the MMP algorithm by Mitchell, Mount and Papadimitriou~\cite{MMP},
the CH algorithm by Chen and Han~\cite{CH} and the ICH algorithm by Xin and Wang~\cite{ICH}. Moreover, a graph-based approach to the vertex-to-vertex all-pairs shortest path problem on a polyhedron is given based on the Saddle Vertex Graph by Ying, Wang, and He~\cite{SVG}.
%%
%Following~\cite{MMP}, there has been extensive studies~\cite{CH,ICH,SVG} on the single-source shortest path problem on a polyhedron in computational geometry and computer graphics.
%%
On the other hand, geodesics on a smooth surface are defined to be locally shortest paths in differential geometry, and have applications in subfields of physics such as mechanics and optics, but relatively less attention has been attracted in the discrete algorithm field. Particularly, the problem of enumerating locally shortest geodesics with given two endpoints on a polyhedron has not been studied to date.

In this paper, firstly we define geodesics on a polyhedron, and formulate the \emph{single-source geodesics enumeration problem}
as a generalization of the single-source shortest path problem on a polyhedron.
%Then, we present a ``naive'' algorithm and an ``efficient'' algorithm, and compare them by computer experiments.
%In this paper, for the first time,
%Then, we consider a generalization of the single-source shortest path problem on a polyhedral surface, called the \textit{single-source geodesics enumeration problem}, defined as follows:
The problem is stated as follows: given a point $s$ on a polyhedral surface and a positive real number $R$, an algorithm is asked to build a data structure $\mathcal{T}$ for queries of enumeration of all (possibly non-shortest) geodesics shorter than $R$, which goes from $s$ to arbitrarily chosen point $t$ on the surface. Such a query is called a \textit{geodesics enumeration query}.
For the purpose, we first introduce a naive version of the data structure, called the \textit{complete geodesic interval tree}, that is computed from a point $s$ on a polyhedral surface and a positive real number $R$ in $O(N\log N)$ time using $O(N)$ space, where $N$ is the number of intervals contained in $\mathcal{T}$.
%However, a difficulty arises when we define such a data structure since there are infinitely many geodesics of length at most $R$ from a fixed $s$. To solve this problem, we use an isomorphism over geodesics introduced by xxx.
%%
Next, we consider the problem of reducing the number of intervals in $\mathcal{T}$, which leads to the reduction of time and space of enumerating geodesic paths.
%For this problem,
To tackle this problem, we present an improved version of the data structure, called the \textit{reduced geodesic interval tree}, where we remove the overlaps between adjacent intervals around a hyperbolic vertex. Moreover, we show that the reduced geodesic interval tree allows a query result to be succinctly encoded as a \textit{single-pair geodesic graph}. By our experiments, it was shown that the reduced geodesic interval tree is smaller than the complete geodesic interval tree, and the reduced geodesic interval tree is better in practical running time and memory consumption.

%\subsection{Related work}

\section{Preliminaries}
Throughout this paper, we simply use the term \emph{polyhedron} (plural \emph{polyhedra}) to mean a (possibly non-convex) polyhedral surface, not solid polyhedron. We deal with simplicial (which means triangulated) orientable polyhedra in $\R^3$, and the symbol $\mathcal P$ is dedicated to mean such one.

\begin{definition}
	Let $v$ be a vertex of $\mathcal P$. Let $\tau$ be the sum of angles around $v$ (measured along the faces) and we call it the \emph{total angle} of $v$. Following~\cite{Straightest}, we say that

	\begin{itemize}
		\item if $\tau < 2 \pi$, $v$ is \emph{spherical;}
		\item if $\tau = 2 \pi$, $v$ is \emph{Euclidean;}
		\item if $\tau > 2 \pi$, $v$ is \emph{hyperbolic.}
	\end{itemize}

	\begin{figure}[h]
		\centering
		\subfloat[spherical]{
			\includegraphics[height=128pt]{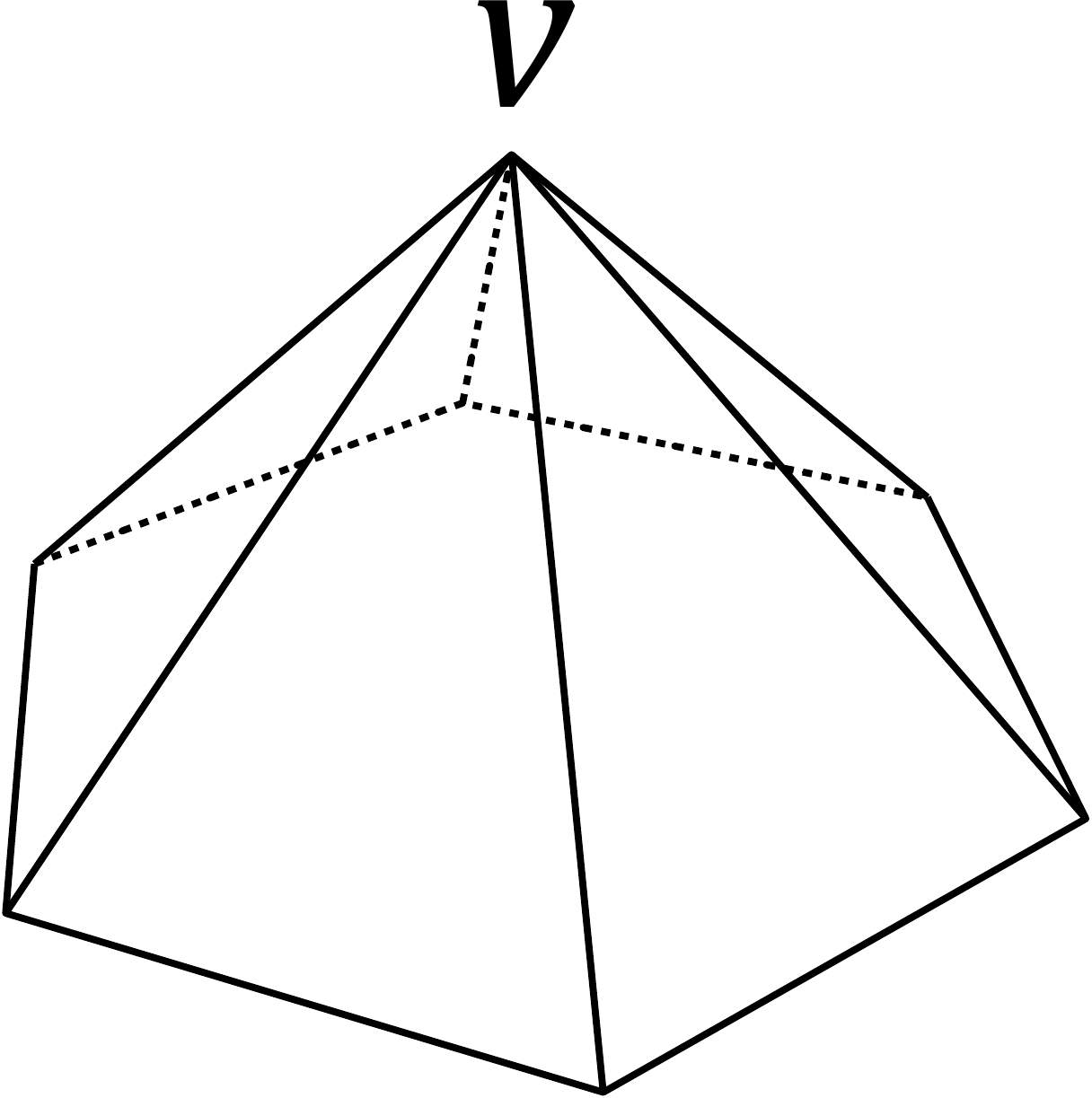}
		}
		\subfloat[hyperbolic]{
			\includegraphics[height=128pt]{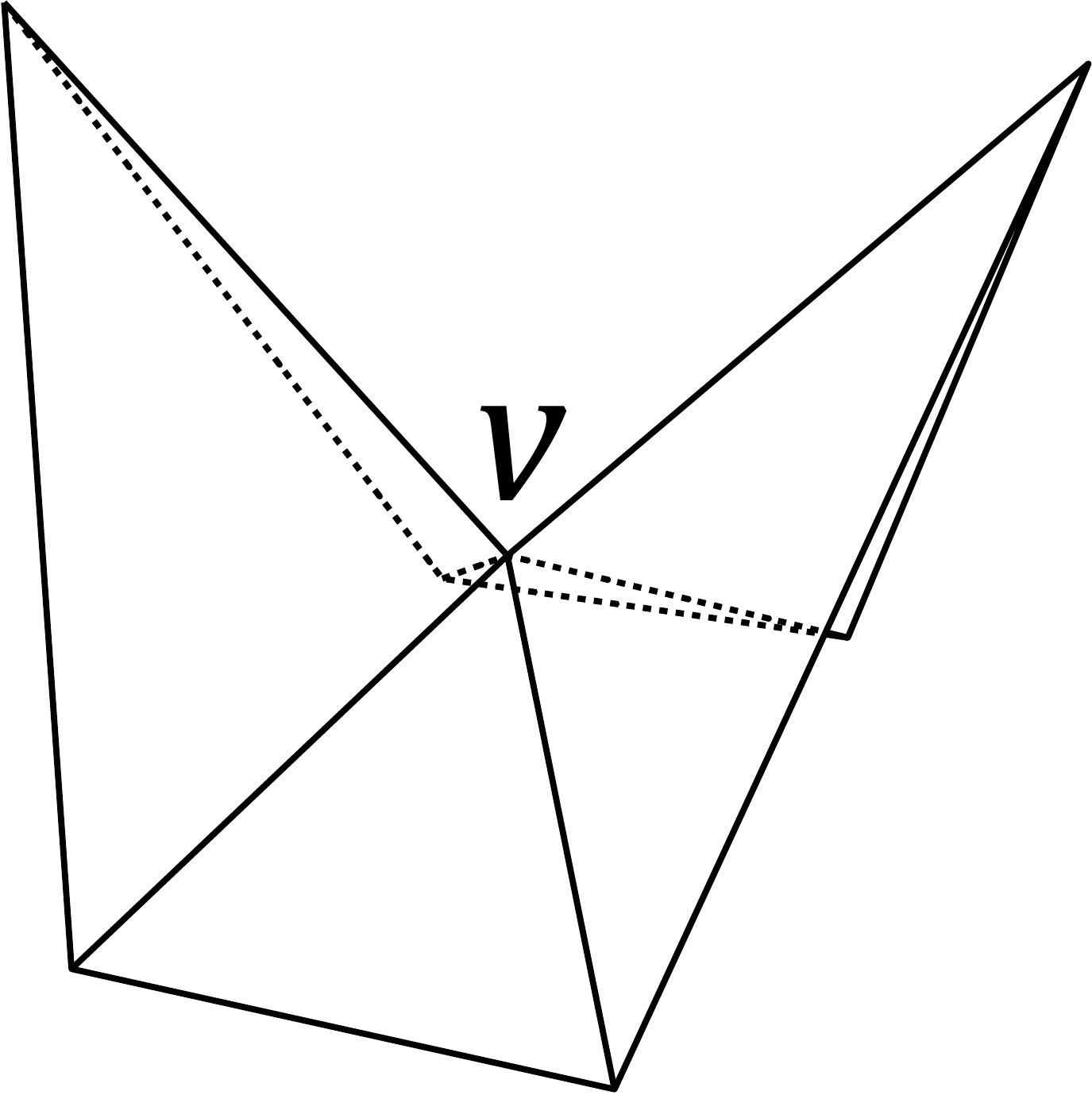}
		}
		\caption{A spherical vertex and a hyperbolic vertex}
	\end{figure}
\end{definition}

\begin{remark}
	If $\mathcal P$ is convex, every vertex is spherical or Euclidean. Note that the angle defect $2\pi - \tau$ can be interpreted as the \textit{discrete Gaussian curvature} at $v$. For example, a discrete analog of the Gauss-Bonnet theorem holds for a polyhedron~\cite{crane2018discrete}.
\end{remark}

We define a geodesic as follows:

\begin{definition} (geodesics)
\label{def_geodesics}
Let $\gamma$ be a path connecting $s$ and $t$ on $\mathcal P$. We say that $\gamma$ is a geodesic if and only if:

\begin{enumerate}
	\item $\gamma$ is straight inside any face, and where $\gamma$ passes through an edge sequence $\mathcal E = (e_1, \dots, e_k)$, $\gamma$ is straight on the unfolding obtained from $\mathcal E$;
	\item where $\gamma$ passes through a vertex, the angle at the vertex made on the both sides of $\gamma$ are greater than or equal to $\pi$ (see Figure~\ref{vertex_figure_0}).
%	Particularly, $\gamma$ never passes through an elliptic vertex.
\end{enumerate}

\begin{figure}[h]
	\centering
	\includegraphics[height=5cm]{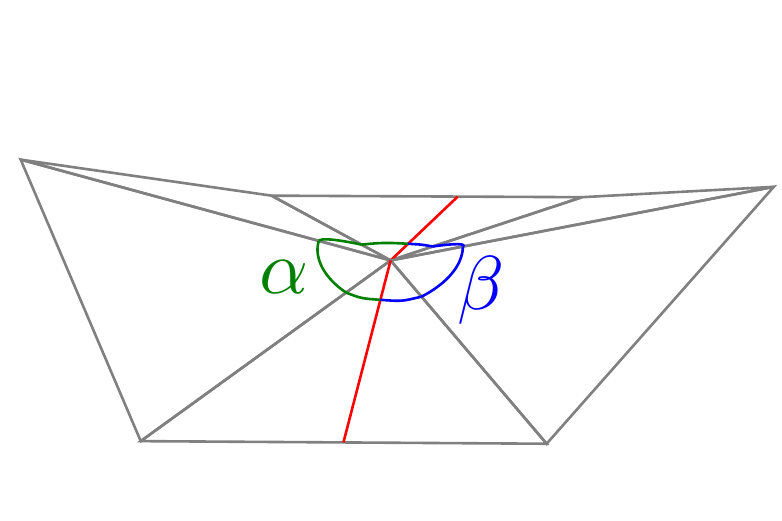}
	\includegraphics[height=5cm]{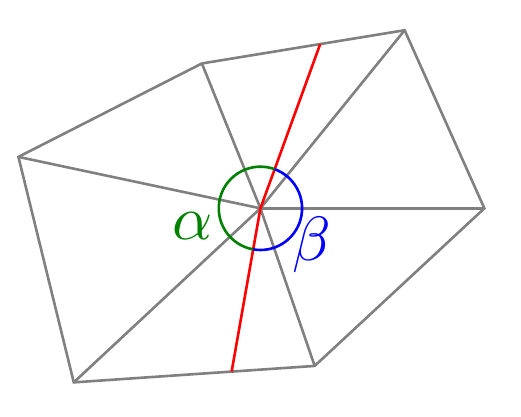}
	\caption{Geodesic passing through a vertex. Here $\alpha \ge \pi$ and $\beta \ge \pi$.\\ Left: 3D perspective view, right: top view}
	\label{vertex_figure_0}
\end{figure}

\end{definition}

\begin{lemma}
	A geodesic never passes through any spherical vertex.
	\begin{proof}
	Since the angle $\alpha$ and $\beta$ at the vertex made on the both sides of $\gamma$ satisfy $\alpha\ge\pi$ and $\beta\ge\pi$ (see Figure~\ref{vertex_figure_0}), the total angle of this vertex is $\tau = \alpha + \beta \ge 2 \pi$.
	\end{proof}
\end{lemma}

\begin{remark}
	Although a geodesic does not pass through a spherical vertex, its endpoints may be spherical vertices. Moreover, a geodesic may pass through an arbitrarily close neighbor of a spherical vertex.
\end{remark}

\begin{remark}
	In Figure \ref{vertex_figure_0}, we depicted the angles at a vertex in two different styles: in a 3D perspective view and a top view. Whenever we present a top-view style figure (like the right one in Figure \ref{vertex_figure_0}), angles described in it are intended to be measured along the faces.
\end{remark}

%Although some authors use the term \textit{geodesic} as a synonym of \textit{shortest path} (for example,~\cite{bose2011survey,SVG}), we gave it a distinct definition, as in differential geometry of smooth surfaces.
%A shortest path is a geodesic, but a geodesic is not necessarily a shortest path. Instead, a geodesic is a locally-shortest path.

Intuitively, a geodesic is a locally-shortest path. Geodesics on a polyhedron, defined above, can be used to approximate geodesics on a smooth surface, defined in differential geometry. Nevertheless, there are some important differences between the discrete geodesics and the smooth geodesics. Let us take a look at some examples.

\begin{figure}%[h]
	\includegraphics[height=7.5cm]{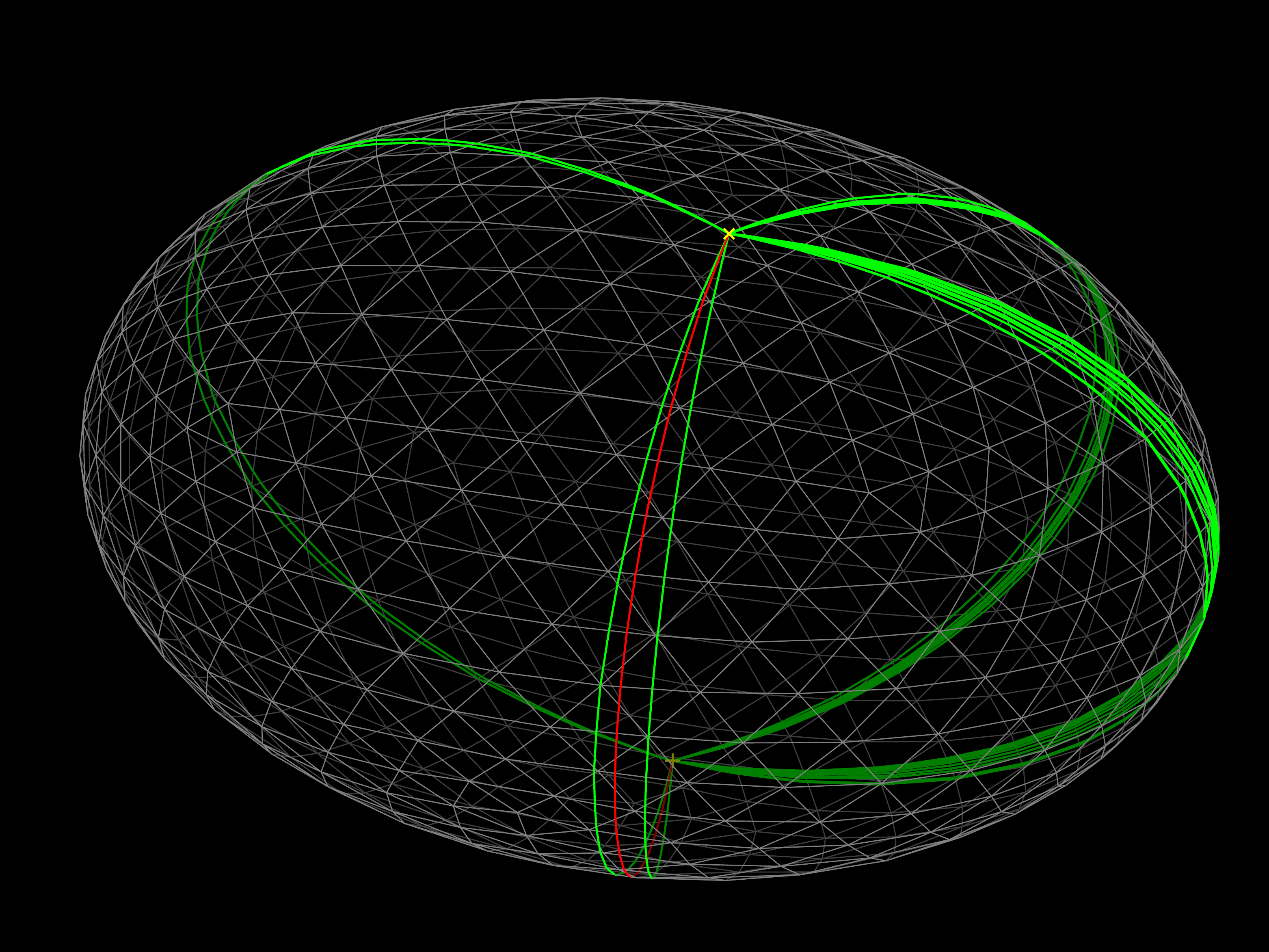}
	\caption{Geodesics on a discrete ellipsoid}
	\label{ellipsoid}
\end{figure}

Figure~\ref{ellipsoid} shows the shortest geodesic (red) and other geodesics (green) on a discrete ellipsoid. Since it is convex (and does not have Euclidean vertices), every vertex is spherical and geodesics pass through no vertices. Instead, there are often multiple similar geodesics that differ in how they ``bypass'' the vertices. As a result, when a smooth surface is discretized into a polyhedron, a single geodesic on the smooth surface often corresponds to multiple geodesics on the discretized polyhedral surface.

\begin{figure}%[h]
	\includegraphics[height=7.5cm]{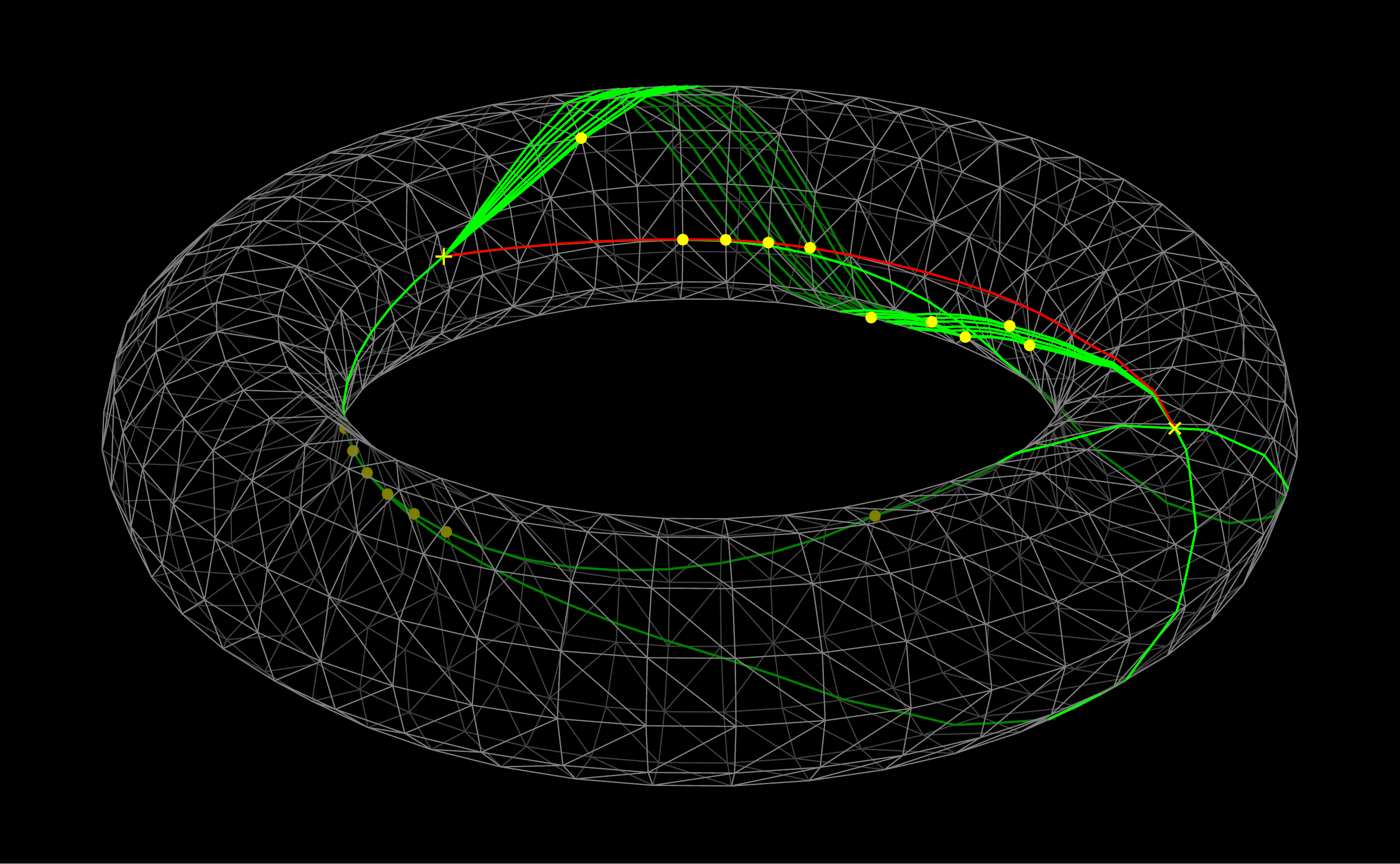}
	\caption{Geodesics on a discrete torus}
	\label{torus}
\end{figure}

If the polyhedron has hyperbolic vertices, the geodesics can pass through them. Figure~\ref{torus} shows that the shortest geodesic (red) passes through consecutive four hyperbolic vertices, as well as some of non-shortest geodesics (green) also pass through several vertices. These vertices are marked in yellow. In general, a geodesic can be decomposed into a sequence of geodesics passing through no hyperbolic vertices. We call them \emph{primitive geodesics}.

\begin{definition}[primitive geodesic]
	A geodesic passing through no hyperbolic vertices is called a \emph{primitive geodesic}.
\end{definition}

\begin{remark}
	Endpoints of a primitive geodesic may be hyperbolic vertices.
\end{remark}

\subsection{Our problem}
Let $\mathcal P$ be a simplicial polyhedron in $\R^3$. When we give a \emph{source} $s$ on $\mathcal P$, we want to compute all geodesics from $s$ to an arbitrary point $t$ on $\mathcal P$. Since we have infinitely many choice of $t$, we consider a query that inputs $t$ and outputs these geodesics. And yet, there are likely to be infinitely many such geodesics and we give an upperbound of their length $R$ to obtain finite result.

%%We remark that even if the lengths of geodesics are bounded from above by $R$, there are huge number of such geodesics. For example, consider a cube with edge length $r$. Then, we have at least $\frac{R}{4r}$ geodesics, which approaches infinity when $R$ grows.

It is widely known that single-source shortest path problems (SSSPs) for graphs or polyhedra can be efficiently solved using a queue or a priority queue. Here, we can think of a generalization of SSSPs, namely, \textit{single-source geodesics enumeration problem} on a polyhedron.

\begin{definition}[single-source geodesics enumeration problem]
\label{problem_def}
Suppose that $\mathcal P$ is a simplicial polyhedron in $\R^3$, $s$ is a point on $\mathcal P$ and $R$ is a positive real number. We define our \emph{single-source geodesics enumeration problem} to be the problem of building data structure that enables to query, for any point $t$ on $\mathcal P$, all geodesics from $s$ to $t$ whose length is less than $R$.
We call an algorithm for this problem a \emph{single-source geodesics enumeration algorithm}, or simply \emph{geodesics enumeration algorithm}.
\end{definition}

\section{Related work}
As far as we know, there is no prior work for enumerating geodesics on a polyhedron. However, there is a variety of research relating to geodesics on polyhedral meshes.

\subsection{Shortest geodesics}
The shortest path problem on a polyhedron has been extensively researched. On a polyhedron without boundary, a shortest path is a geodesic in our sense~\cite{MMP}, thus the shortest path problem is equivalent to the shortest geodesic problem. Shortest geodesic algorithms can be divided into exact algorithms and approximate algorithms. Since the interest of this paper is exact algorithms, approximate algorithms are not discussed in detail here. Detailed survey of this topic is given by~\cite{bose2011survey,crane2020survey}.

\subsubsection{Interval propagation algorithms for the SSSP}
The \emph{MMP algorithm} given by Mitchell, Mount and Papadimitriou~\cite{MMP} is an important exact algorithm for the SSSP on a polyhedron. It retains \emph{intervals} on each edge, so that the shortest path to any point within an interval has the same combinatorial structure (faces, edges and vertices). Information required to reconstruct geodesics is appended to these intervals. However, each interval only represents geodesics passing through a particular face among the two faces incident to the edge. This can be interpreted that each interval is assigned to a directed edge and only represents geodesics through one side (in the original paper, it is the right side along the directed edge).

In the earliest stage of the MMP algorithm, intervals are created only for the edges facing $s$. Each interval is \emph{propagated} using the \emph{continuous Dijkstra} scheme. When it detects a shortest geodesic reaching a hyperbolic vertex, it generates the intervals representing the geodesics passing through that vertex. Intervals on the same directed edge are ordered, and a newly-propagated interval is inserted into the ordered list of the intervals already-existing on the directed edge. Then, it performs \emph{trimming} among the new interval and adjacent intervals. By trimming, intervals are cut to ensure shortestness against other intervals, and intervals on a directed edge become disjoint. Intervals may become empty as a result of trimming, and such intervals are not propagated. When no non-empty intervals are newly created and the priority queue becomes empty, the algorithm terminates and outputs the shortest geodesics. Its computational complexity given by the original authors is $O(n^2 \log n)$ time and $O(n^2)$ space, where $n$ is the number of the edges of the input polyhedron. However, according to an experiment by Surazhsky et al.~\cite{Surazhsky}, its practical complexity is subquadratic and could be considered as approximately $O(n^{1.5} \log n)$ time and $O(n^{1.5})$ space. They analyzed the reason behind it as that, the number of intervals per edge is approximately $O(n^{0.5})$ in practice, despite the $O(n)$ estimation by MMP.

Most of exact algorithms for the SSSP on a polyhedron, published after the MMP algorithm, contain the concept of interval propagation. The \emph{CH algorithm} by Chen and Han~\cite{CH} uses a FIFO queue instead of a priority queue. Its theoretical complexity is $O(n^2)$ time and $O(n)$ space, but its practical performance is not as good as the MMP algorithm. The \emph{ICH} (improved CH) \emph{algorithm} by Xin and Wang~\cite{ICH} introduces into the CH algorithm a priority queue as well as several new rules to exclude intervals not contributing to any shortest paths. In theory, the usage of a priority queue increases its time complexity to $O(n^2 \log n)$, but greatly improves its practical performance. While the MMP algorithm requires inserting an interval into an ordered list and solving a quadratic equation to get the intersection of the interval and certain hyperbola, the ICH algorithm does not. As a result, the ICH algorithm may outperform the MMP algorithm despite the ICH algorithm may generate more intervals than the MMP algorithm. Moreover, the ICH algorithm does not require the history of the propagated intervals to be retained. As a result, the space complexity of the ICH algorithm is $O(n)$, which is significantly smaller than the MMP algorithm.

\subsubsection{Saddle Vertex Graph}
The \emph{Saddle Vertex Graph} (SVG) by Ying, Wang and He~\cite{SVG} is an approach to the vertex-to-vertex all-pairs shortest path problem on a polyhedron. They noticed that a shortest geodesic between two hyperbolic vertices may be shared by multiple longer shortest geodesics, and reduced the problem to the well-known shortest path problem on a graph by precomputing shortest primitive geodesics connecting two vertices. However, when the given polyhedron is convex, it has no hyperbolic vertices and there is no difference from precomputing the shortest path for every pair of vertices. However, they observed real-world meshes contain 40-60\% hyperbolic vertices and more than 90\% of shortest paths pass through at least one hyperbolic vertex. Also, they considered the exact SVG is too large to be tractable for large meshes and proposed a method to sparsify it, and evaluated error by computer experiments.

\subsection{General geodesics}
Geodesics, which are not limited to shortest, also have been researched in the context of geodesic tracing and path shortening.

\subsubsection{Geodesic tracing}
A classical problem about a smooth surface in differential geometry is to trace the smooth geodesic from a given point $p$ and a tangent direction $v$. Concerning the corresponding problem on a polyhedron, this can be done at almost every point. However, a geodesic cannot proceed beyond a spherical vertex, and it cannot uniquely determine its direction when it hits a hyperbolic vertex. To cope with this problem, Polthier and Schmies suggested a \emph{straightest geodesic} which goes through a vertex to halve the total angle of the vertex~\cite{Straightest}. However, it is not necessarily locally-shortest anymore, and does not have continuity respect to $p$ and $v$, i.e. it jumps when it moves onto or across a non-Euclidean vertex, thus its nature vastly differs from a geodesic on a smooth surface. To fill this gap, Cheng, Miao, Liu, Tu and He suggested a method to trace smooth geodesics on a tangent-continuous curved surface constructed from the input polyhedron using PN-triangles, and evaluated its accuracy by computer experiments~\cite{IVP}.

\subsubsection{Path shortening}
In this approach, an input polyline on a polyhedron is shortened until it becomes a geodesic. The shortening process is performed according to local configuration of the path. It is implemented in e.g.~\cite{CyberTape,martinez2005,xin2007} and they used it to refine a path obtained by Dijkstra's algorithm on the edge graph or other approximate shortest path algorithms on the polyhedron.

Recently, a flip-based algorithm was developed by Sharp and Crane~\cite{Flip}. An initial path is given as an edge sequence, and until the path becomes a geodesic, the algorithm modifies the triangulation by flipping an edge so that the new shortened path is still on the edges of the new triangulation. It works on \emph{intrinsic triangulations} of a polyhedron --- that is, the new triangulation is intrinsically embedded in the original surface, and its edges are no longer straight when it is viewed as an object in $\R^3$. It allows the geometry of the original surface to remain unchanged by flipping. Their method can yield not only open geodesics but also closed geodesics. Although they did not give worst-case bounds, they proved that it always obtains a geodesic by finite operations, and observed that its practical running time is on the order of milliseconds.

\section{Geodesics and intervals}
\begin{figure}
	\includegraphics[height=10cm]{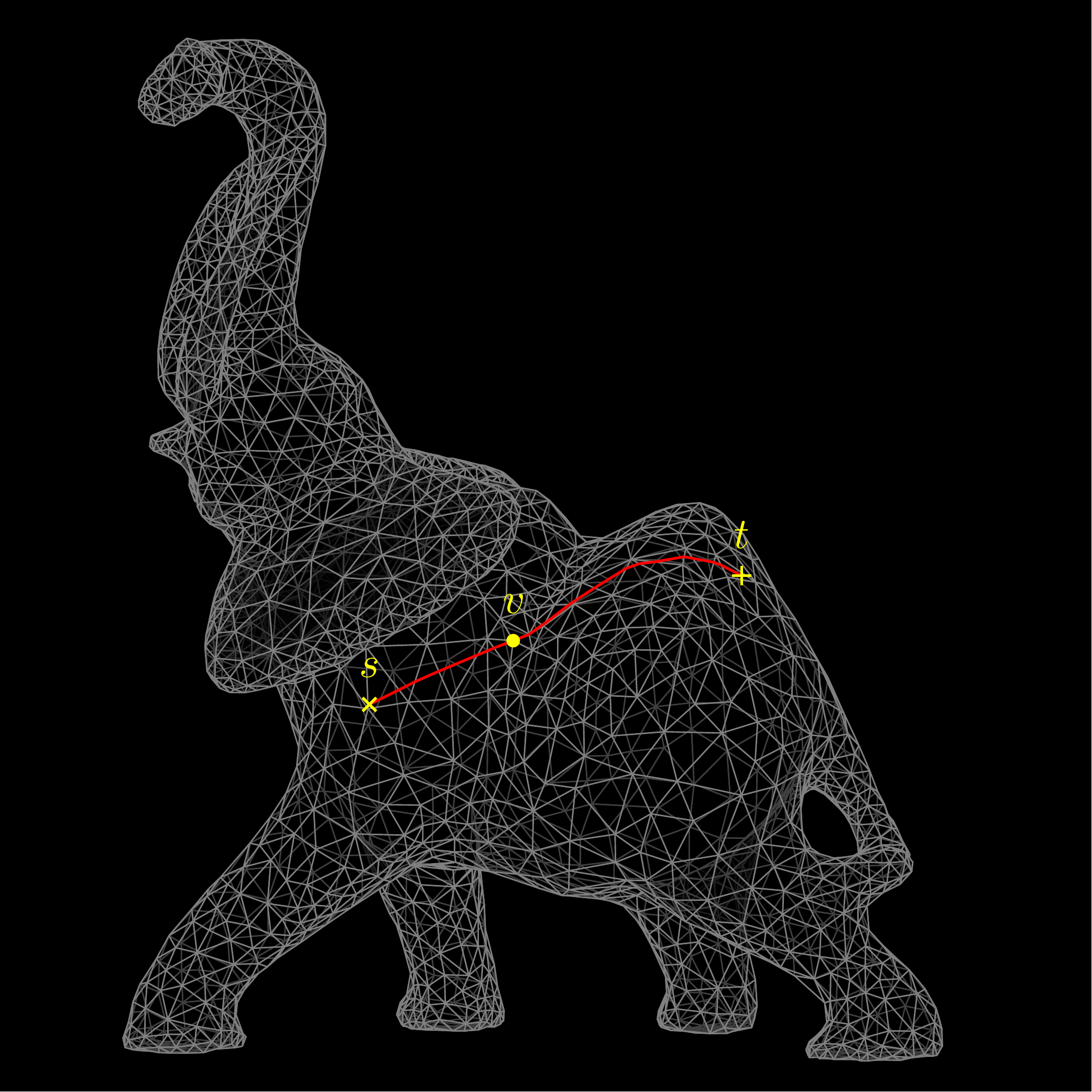}
	\caption{A geodesic from $s$ to $t$ via $v$}
	\label{elephant_shortest}
\end{figure}

\begin{figure}
	\includegraphics[height=6cm]{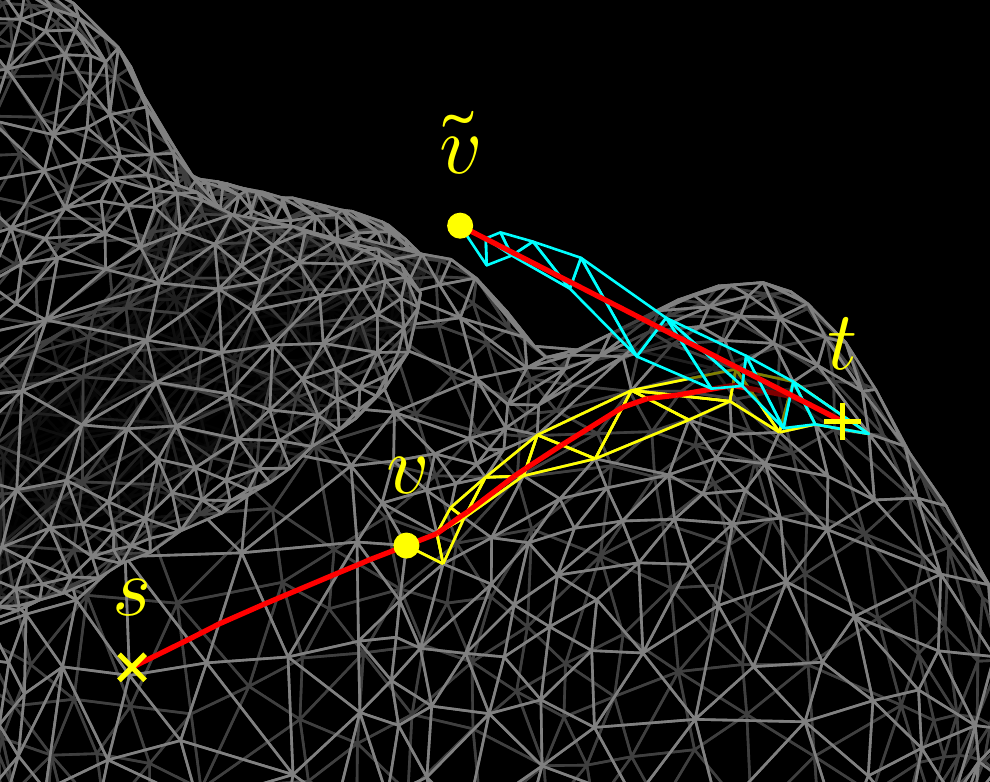}
	\caption{Geodesic and unfolding}
	\label{elephant_unfolding_zoomed}
\end{figure}

To explain how intervals can be used to express geodesics, we begin with an observation of geodesics on a polyhedron. In Figure~\ref{elephant_shortest}, a geodesic $g$ from the source $s$, to $t$ on a face $\sigma$, is shown in red. Since $g$ passes through a vertex $v$, it can be decomposed into the two primitive geodesics, between $sv$ and between $vt$. By Definition~\ref{def_geodesics} (1), a primitive geodesic can be unfolded into a line segment. Particularly, the primitive geodesic between $vt$ can be unfolded into the line segment $\tilde vt$ on the plane containing $\sigma$ (Figure~\ref{elephant_unfolding_zoomed}).

\begin{figure}
	\includegraphics[height=3cm]{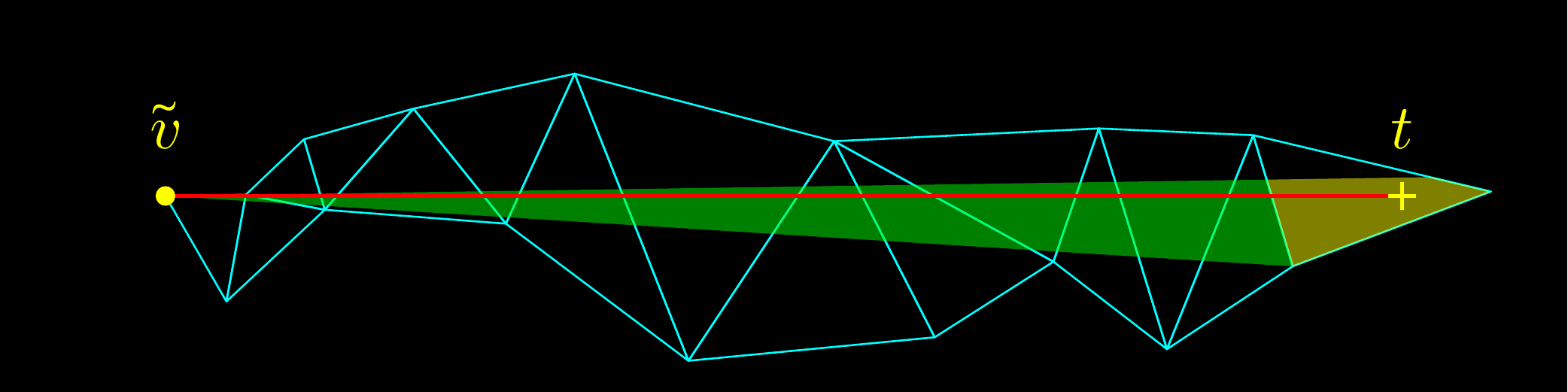}
	\caption{The same unfolding in 2D}
	\label{elephant_unfolding_2d}
\end{figure}

This unfolding of the geodesic between $vt$ is shown in 2D in Figure~\ref{elephant_unfolding_2d}. As a simple observation, when one moves $t$ in the region shaded in yellow, the line segment $\tilde vt$ still gives a geodesic on the unfolding. This region is chosen so that the line segment $\tilde vt$ does not go out of the unfolding and satisfies the condition (2) in Definition~\ref{def_geodesics} at $v$.

\begin{figure}
	\includegraphics[height=3cm]{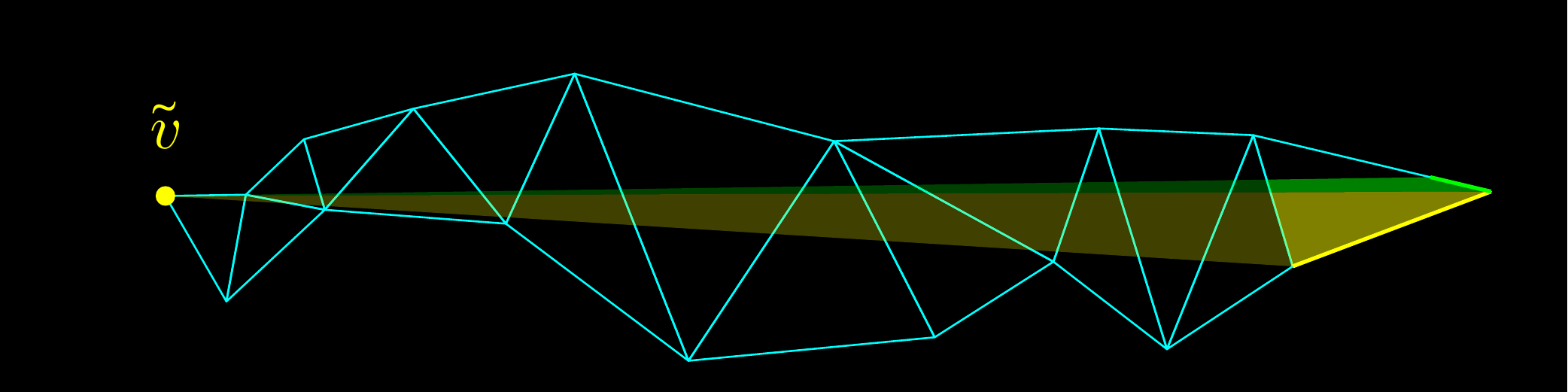}
	\caption{Geodesic and intervals}
	\label{elephant_unfolding_2d_2}
\end{figure}

When $t$ is inside a face, one can extend the geodesic until hitting an edge, thus here we only consider geodesics to a point on an edge. We introduce concept of intervals generalizing the ones in the MMP algorithm (or other interval propagation algorithm) to express geodesics to a particular range on an edge. Figure~\ref{elephant_unfolding_2d_2} shows an outline of this expression. The yellow and green line segments represent the range of the intervals and indicate the range on which geodesics can be given in this unfolding. The region darkly shaded in their respective color indicates the region in which the interval is used in the geodesic query. As a remark, when a geodesic passes through no vertices, we consider the unfolding of the whole geodesic from $s$, and when it passes through multiple vertices, we consider the unfolding of the primitive geodesic between the last vertex and $t$.

\begin{figure}
	\centering
	\includegraphics[height=5cm, pagebox=cropbox]{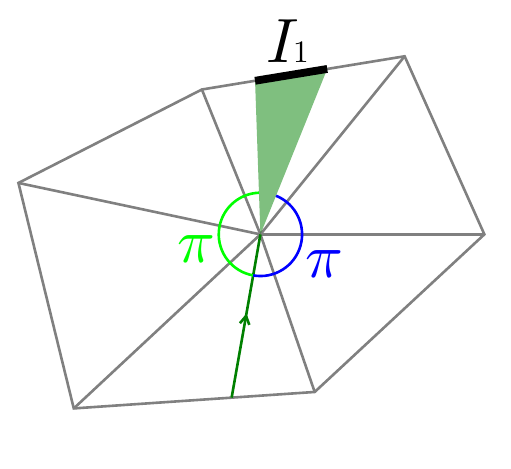}
	\includegraphics[height=5cm, pagebox=cropbox]{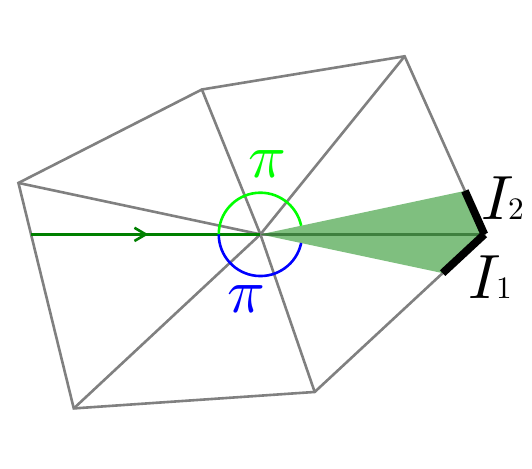}
	\caption{pseudo-source intervals, made at a hyperbolic vertex}
	\label{vertex_figures}
\end{figure}

Figure~\ref{vertex_figures} shows the intervals to express the geodesics immediately after passing through a vertex. We call them \emph{pseudo-source intervals}.

\begin{definition}
	\label{interval}
	An interval $I$ is defined to be the following data structure:
	\begin{itemize}
		\item $I.\Parent$: the interval which generated $I$ (by propagation)
		\item $I.\Edge$: the target edge
		\item $I.\Face$: the target face
		\item $I.\Extent$: the target line segment on $I.\Edge$ (identified with $I$ itself)
		\item $I.\Center$: the unfolded position of the last vertex
		\item $I.\Depth$: the length of the geodesic between $s$ and the last vertex
	\end{itemize}
	It is used to express all geodesics reaching the line segment $I.\Extent$ on the edge $I.\Edge$ through the last vertex. If the geodesics have not passed through a hyperbolic vertex yet, $I.\Center$ is the unfolded position of $s$ and $I.\Depth$ is 0.

	We regard $I$ to be oriented so that $I.\Face$ is seen on the left side along $I.\Edge$. We define the starting point of $I$ with respect to this orientation. Moreover, $I$ has the following two functions:
	\begin{itemize}
		\item $I.\Project(p : p \in I.\Face)$ := the intersection point of $I.\Edge$ and the line connecting $p$ and $I.\Center$
		\item $I.\Distance(p : p \in I.\Face)$ := (the distance of $p$ and $I.\Center$) + $I.\Depth$
	\end{itemize}

	The interval $I$ can yield a geodesic at $p \in I.\Face$ if $I.\Project(p) \in I.\Extent$, and its length is given by $I.\Distance(p)$.
\end{definition}

\begin{remark}
	$I.\Center$ can be expressed in either the 3D global coordinate system or a 2D local coordinate system on $I.\Face$. 2D local coordinates are slightly faster and memory efficient, as well as ensure that given coordinates are always on the plane. Our implementation uses the 3D global coordinate system in input and output, but uses a 2D local coordinate system in internal storage and computation, and converts one to the other when necessary. That means, our algorithms could run on a polyhedron in higher-dimensional Euclidean space as well.
\end{remark}

\begin{remark}
	Our algorithms can also work on a self-intersecting polyhedral surface. In this case, geodesics are not bent at the intersection and pass through it as if there were no intersection --- geodesics are completely defined locally. On the other hand, we assume the orientability of the surface, which may not be the case for a self-intersecting (or higher-dimensional) polyhedron. If this is an issue, one can take the orientable double covering of the surface, as demonstrated in Figure~\ref{fig_roman}. In this figure, geodesics are computed on a discretized version of the orientable double covering, which is homeomorphic to the sphere, of the (non-orientable) Roman surface (a realization of the real projective plane in $\R^3$). The source $s$ is given once, but the target $t$ is given twice, correspondingly to its two images on the double covering. In the figure, two different colors are used accordingly.
	\begin{figure}
		\centering
		\includegraphics[height=8cm]{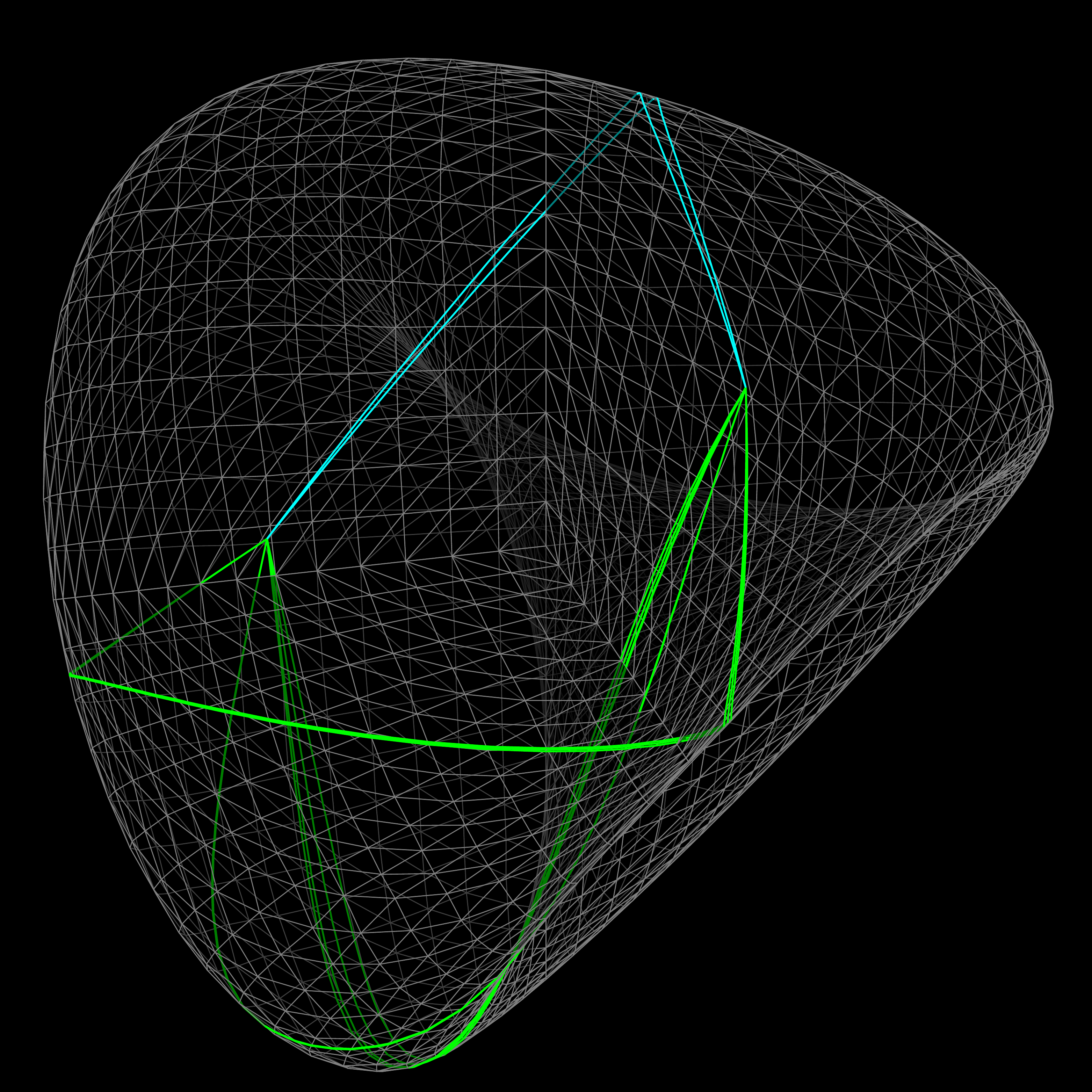}
		\caption{Geodesics on a discretized Roman surface}
		\label{fig_roman}
	\end{figure}
\end{remark}

\begin{remark}
Our implementation uses the half-edge data structure as the internal representation of a polyhedron, thus every edge is directed. Here the 2D local coordinate system is defined by the directed edge $e = I.\Edge$ so that its origin is the starting point of $e$, its positive $x$ direction is the direction of $e$, its $y$ axis is orthogonal to the $x$ axis and any point inside $I.\Face$ has positive $y$ value. Complex numbers are convenient for expressing the local coordinates, because
\begin{itemize}
	\item we often need an 1D parametric coordinate on $e$ as well. Since a real number is considered to be a complex number whose imaginary part is zero, we can naturally treat it as a special case of 2D local coordinates;
	\item when we express $I.\Center$ in the 2D local coordinate system, we need to apply coordinate transformation to propagate an interval. This transformation consists of 2D rotation and translation, which can be expressed in terms of complex multiplication and addition;
	\item functions such as $\mathrm{abs}$ and $\mathrm{arg}$ are also useful to implement our algorithm.
\end{itemize}
\end{remark}

\section{Naive geodesics enumeration algorithm}
We can make a naive geodesics enumeration algorithm by propagating intervals without the trimming process in the MMP algorithm. All intervals generated in this way form a tree structure by the \emph{Parent} pointer. we call it the \emph{complete geodesic interval tree} or the \emph{complete GIT}.

This algorithm firstly performs initialization of generating several intervals, and proceeds by processing events. We define \emph{propagation} to be the act of generating one or more new intervals by processing an event. Events consist of the following two types:

\begin{itemize}
	\item edge event : when a geodesic given by $I$ reaches $I.\Extent$ for the first time
	\begin{itemize}
		\item this event is associated with $I$
		\item time of occurrence: (the distance of $I.\Extent$ and $I.\Center$) + $I.\Depth$
	\end{itemize}
	\item vertex event : when a geodesic reaches a vertex $v$
	\begin{itemize}
		\item this event is associated with the interval $I$ of which the starting point is $v$
		\item time of occurrence: (the distance of $v$ and $I.\Center$) + $I.\Depth$
	\end{itemize}
\end{itemize}

We introduce an \emph{event queue} to manage the order of events. It is a priority queue, and lets the algorithm process events in the ascending order of their time of occurrence.

The outline is described in Algorithm~\ref{algo_main} as a pseudocode.
\begin{algorithm}%[h]
	\caption{(Building the geodesic interval tree)}
	\label{algo_main}
	\begin{algorithmic}[1]
	\Function {BuildGeodesicIntervalTree} {$s$: source, $R$: upperbound of length}
		\State $Q$ := the event queue
		\State \Call {Initialize}{$Q$, $s$}
		\While {the time of occurence of the top event of $Q$ $< R$}
			\State $Q$.\textsc{Pop}().\textsc{Handle}($Q$)
		\EndWhile
	\EndFunction
	\end{algorithmic}
\end{algorithm}

\begin{remark}
	In Algorithm 1 (and Definition \ref{problem_def}), we assume that $R$ is given ahead of time. Alternatively, we can rewrite the line 4 with another arbitrary terminating condition (or as an infinite loop which can be stopped by a human), and when the loop terminates, $R$ can be obtained as the time of occurrence of the top event of $Q$. This also applies to the improved algorithm in the next section.
\end{remark}

\subsection{Initialization}
\begin{figure}%[h]
	\centering
	\includegraphics[height=3cm, pagebox=cropbox]{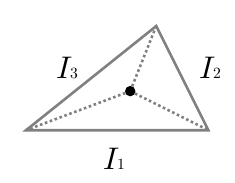}
	\includegraphics[height=3cm, pagebox=cropbox]{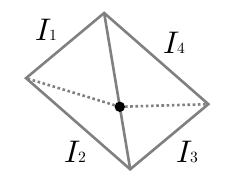}
	\includegraphics[height=3cm, pagebox=cropbox]{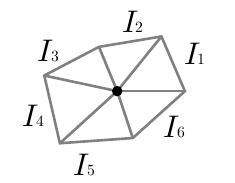}
	\caption{Initialization}
	\label{initialization}
\end{figure}

Firstly, the intervals for all edges facing $s$ are generated. Figure~\ref{initialization} shows the cases of $s$ being inside a face (left), inside an edge (middle), and at a vertex (right). For each $I$ of these initial intervals, $I.\Parent$ is Null, $I.\Center$ is $s$, $I.\Depth$ is 0, $I.\Extent$ is the whole $I.\Edge$, and $I.\Face$ is the face containing both $s$ and $I.\Edge$. For each $I$, the associated edge event and vertex event are inserted into the event queue. (Algorithm~\ref{algo_init})

\begin{algorithm}%[h]
	\caption{(Initialization)}
	\label{algo_init}
	\begin{algorithmic}[1]
	\Function {Initialize} {$Q$, $s$}
		\State $I_1 \dots I_k$ := the initial intervals
		\For {$i$ in $1 \dots k$}
			\State $Q$.\Call{Push}{EdgeEvent($I_i$)}
			\State $Q$.\Call{Push}{VertexEvent($I_i$)}
		\EndFor
	\EndFunction
	\end{algorithmic}
\end{algorithm}

\subsection{Procedure for edge events}
\begin{figure}%[h]
	\centering
	\includegraphics[height=5cm, pagebox=cropbox]{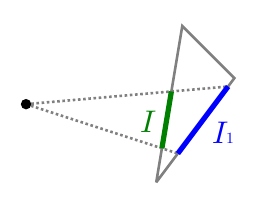}
	\includegraphics[height=5cm, pagebox=cropbox]{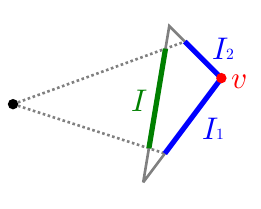}
	\caption{Projection of an interval}
	\label{propagation}
\end{figure}

When an edge event occurs, the associated interval $I$ is projected from $I.\Center$ into the opposite edges of $I.\Edge$ (Figure~\ref{propagation}). The projection result is on either one edge (left) or two edges (right). For each newly created interval $I_i$ (left: $I_1$, right: $I_1, I_2$), $I_i.\Parent = I$, $I_i.\Depth = I.\Depth$ and $I_i.\Face$ is the triangular face in Figure~\ref{propagation}. $I_i.\Center$ is obtained by certain 3D rotation around $I.\Edge$ to make it coplanar with $I_i.\Face$. For each $I_i$, the associated edge event is inserted into the event queue. In the two-interval case (right), the vertex event at the intermediate vertex $v$ is associated with the interval starting at $v$ (that is $I_2$) and inserted into the event queue. (Algorithm~\ref{algo_edge_event})

\begin{algorithm}%[h]
	\caption{(Processing an edge event)}
	\label{algo_edge_event}
	\begin{algorithmic}[1]
	\Function {EdgeEvent.Handle} {$Q$}
		\State $I$ := the associated interval
		% \State Record $I$ on $I.\Edge$
		\If {$I$ is projected into one interval $I_1$}
			\State $Q$.\Call{Push}{EdgeEvent($I_1$)}
		\Else \hspace{1pt} // $I$ is projected into two intervals $I_1$ and $I_2$
			\State $Q$.\Call{Push}{EdgeEvent($I_1$)}
			\State $Q$.\Call{Push}{EdgeEvent($I_2$)}
			\State $Q$.\Call{Push}{VertexEvent($I_2$)}
		\EndIf
	\EndFunction
	\end{algorithmic}
\end{algorithm}

\subsection{Procedure for vertex events}
When a vertex event occurs, the associated interval $I$ is recorded on $v$ as it gives the geodesic arriving at $v$.
% If $v$ is not hyperbolic, simply $I$ is recorded on $v$ and nothing actually happens.
If $v$ is hyperbolic, one or more pseudo-source intervals are generated. In Figure~\ref{vertex_figures}, one interval in the left subfigure, two intervals in the right are generated. For each newly-created interval $I_i$, $I_i.\Parent = I$, $I_i.\Center = v$, $I_i.\Depth$ is the length of the geodesic to $v$ (= the time of occurrence of this event), and the associated edge event and vertex event (if exists) are inserted into the event queue. (Algorithm~\ref{algo_vertex_event_naive})

\begin{algorithm}%[h]
	\caption{(Processing a vertex event, for the complete GIT)}
	\label{algo_vertex_event_naive}
	\begin{algorithmic}[1]
	\Function {VertexEvent.Handle} {$Q$}
		\State $I$ := the associated interval
		\State $v$ := the starting point of $I$
		\State Record $I$ on $v$
		\If {$v$ is not hyperbolic}
			\State \Return
		\EndIf
		\State $I_1 \dots I_k :=$ the corresponding pseudo-source intervals
		\For {$i$ in $1 \dots k$}
			\State $Q$.\Call{Push}{EdgeEvent($I_i$)}
		\EndFor
		\For {$i$ in $2 \dots k$}
			\State $Q$.\Call{Push}{VertexEvent($I_i$)}
		\EndFor
	\EndFunction
	\end{algorithmic}
\end{algorithm}

% \textbf{A Trial}:
% Given a starting point $s$, two geodesics $g, g'$ starting from $s$ and ends at some face $F\in P$ are \textit{isomorphic}, denoted $g \equiv_{s,F} g'$, if they intersect the same sequence of edges and saddle vertices.
% %Formally, for any geodesic $g$ from $s$ to $F$, we denote by $\sigma(g) = (o_1, \dots, o_\ell)$ be the sequence of
% Let $\equiv_s := \bigcup_{F\in P} \equiv_{s,F}$ be the union of equivalence relations for all faces.
% We observe that each branch from the root to some node labeled with face $F$ in the interval tree $\mathcal{T}$ contains the set of all isomorphic geodesics $\{ g' \mid g \equiv_{s,F} g' \}$ to some geodesics represented by the branch.

\subsection{Geodesics enumeration query}
After building the complete geodesic interval tree, it can accept the geodesics enumeration query which inputs a point $t$ on $\mathcal P$ and outputs the set $G_{st}^R$ of geodesics from $s$ to $t$, whose length is less than $R$.
% First, using Algorithm~\ref{algo_query_intervals}, the set of intervals yielding geodesics of $G_{st}^R$ is obtained.
First, we must determine which intervals are responsible for the output. It can be done using the \textsc{GetIntervals} function:

\begin{definition}
	\label{def_get_intervals}
	We define the \textsc{GetIntervals}($t$) function as follows:
	\begin{itemize}
		\item If $t$ is a vertex $v$, \textsc{GetIntervals}($t$) returns the set of intervals recorded on $v$.
		\item If $t$ is on an edge $e$, \textsc{GetIntervals}($t$) returns the set of all generated intervals $I$ such that $I.\Edge = e$, $t \in I.\Extent$ and $I.\Distance(t) < R$.
		\item If $t$ is on a face $f$, \textsc{GetIntervals}($t$) returns the set of all generated intervals $I$ such that $I.\Face = f$, $I.\Project(t) \in I.\Extent$ and $I.\Distance(t) < R$.
	\end{itemize}
\end{definition}

For each obtained interval, using Algorithm~\ref{algo_construct_geodesic}, the geodesic is constructed from $t$ to $s$ by backtracking $I.\Parent$. The whole procedure is given by Algorithm~\ref{algo_geodesic_query_naive}. In the pseudocode, \textsc{Intersect} is the function of getting the intersection, and \textsc{AddFront} is the operation of inserting a point at the front of the list.

\begin{algorithm}%[h]
	\caption{(Construct a geodesic)} \label{algo_construct_geodesic}
	\begin{algorithmic}[1]
	\Function {ConstructGeodesic} {$I$, $p$}
		\State $g$ := $(p)$
		\While {$I.\Parent \ne \mbox{Null}$}
			\State $e$ := $I.\Parent.\Edge$
			\State $p$ := \Call{Intersect}{$e$, the line segment connecting $p$ and $I.\Center$}
			\State $g$.\Call{AddFront}{$p$}
			\State $I$ := $I.\Parent$
		\EndWhile
		\State $g$.\Call{AddFront}{$s$}
		\State \Return {$g$}
	\EndFunction
	\end{algorithmic}
\end{algorithm}

\begin{algorithm}%[h]
	\caption{(Geodesics enumeration query, for the complete GIT)} \label{algo_geodesic_query_naive}
	\begin{algorithmic}[1]
	\Function {GeodesicEnumQuery} {$t$}
		\State $G$ := $\emptyset$ (the set of geodesics for output)
		\For {$I$ in \Call{GetIntervals}{$t$}} %\Comment Algorithm~\ref{algo_query_intervals}
			\State $G$.\Call{Add}{\textsc{ConstructGeodesic}($I$, $t$)} \Comment Algorithm~\ref{algo_construct_geodesic}
		\EndFor
		\State \Return {$G$}
	\EndFunction
	\end{algorithmic}
\end{algorithm}

\section{Improved geodesics enumeration algorithm}
\begin{figure}
	\centering
	\includegraphics[height=8cm]{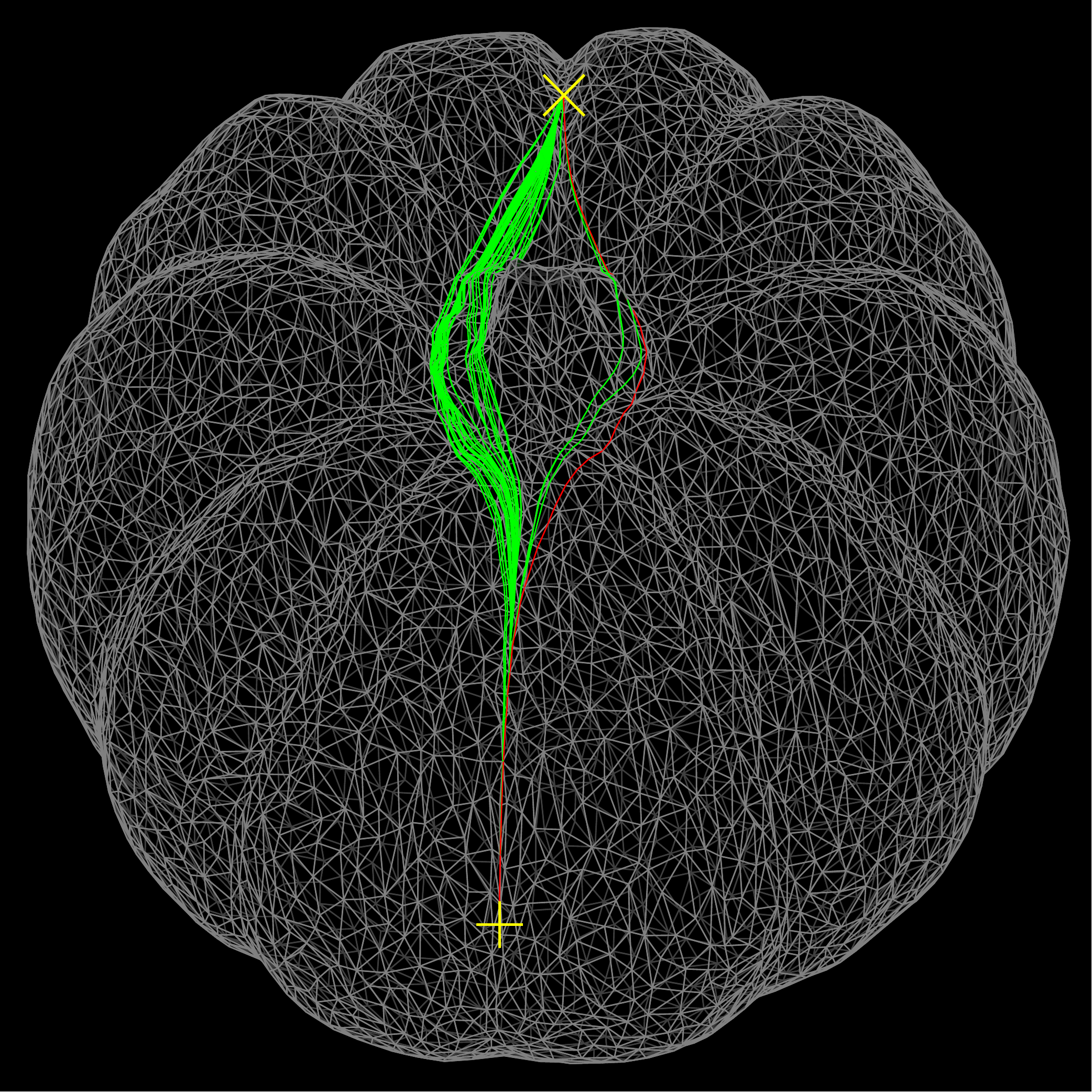}
	\caption{Geodesics on a pumpkin}
	\label{pumpkin}
\end{figure}

\begin{figure}
	\centering
	\includegraphics[height=7cm, pagebox=cropbox]{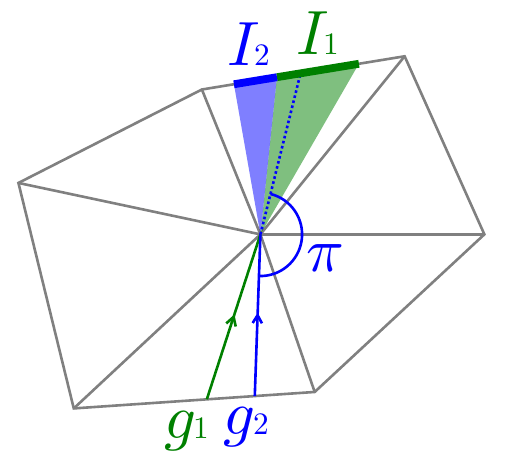}
	\caption{When $g_2$ arrives at a hyperbolic vertex after $g_1$, $I_2.\Extent$ is chosen not to have overlap with $I_1.\Extent$.}
	\label{vertex_figure_overlap}
\end{figure}

The difference between the MMP algorithm and the geodesics enumeration algorithm in the previous section based on the complete geodesic interval tree is that, since we are also interested in non-shortest geodesics, our intervals are never trimmed. However, this change largely increases the computational complexity. Here, when $\mathcal P$ is non-convex, we can reduce required amount of time and memory by placing pseudo-source intervals without overlap. We can understand the redundancy of the naive algorithm using Figure~\ref{pumpkin}. In the figure, the source $s$ is indicated as the yellow cross sign ($\times$) and the target $t$ is indicated as the yellow plus sign ($+$). While multiple geodesics are outgoing from $s$, they merge at some hyperbolic vertices until reaching $t$ and they all share some part near $t$. However, the naive algorithm cannot recognize and utilize this property; the shared part of the geodesics is independently encoded in multiple intervals and rediscovered for each geodesic during the query process. The basic idea of improvement is shown in Figure~\ref{vertex_figure_overlap}. In this figure, a geodesic $g_1$ already arrived at a hyperbolic vertex and yielded a pseudo-source interval $I_1$. Now, another geodesic $g_2$ arrives at the vertex. Although $g_2$ can go through the dotted blue line, $I_2$ can be chosen to exclude the region already searched by $I_1$. As we will see later, we can still restore all geodesics in the query process.

\subsection{Procedure of reduction}
\begin{figure}
	\centering
	\includegraphics[height=7cm, pagebox=cropbox]{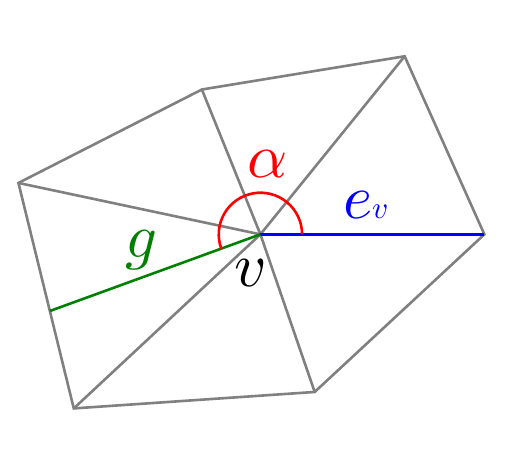}
	\caption{Fix $e_v$ and encode the direction of $g$ into $\alpha$}
	\label{vertex_figure_with_e_v}
\end{figure}

For the purpose of generalizing this argument, for each hyperbolic vertex $v$, we arbitrarily choose an edge $e_v$ among those incident to $v$ and we fix the choice (Figure~\ref{vertex_figure_with_e_v}). It allows us to numericalize the direction of $g$ (seen from $v$) into $\alpha$ (measured along the faces). When $g$ is incoming into $v$, we call $\alpha$ the \emph{incoming angle} of $g$, and when $g$ is outgoing from $v$, we call $\alpha$ the \emph{outgoing angle} of $g$. Also, we use the term \emph{outgoing angle range} $\iota = [\mu, \nu]$ ($\nu - \mu < \tau$) to assert that all values within this range are considered as outgoing angles, here $\tau$ is the total angle of $v$. By convention, if $\mu > \nu$ then $\iota$ is empty, and if $\nu \ge \tau$ then all values within $\iota$ are subject to the ``mod $\tau$'' operation.

\begin{figure}
	\centering
	\includegraphics[height=7cm, pagebox=cropbox]{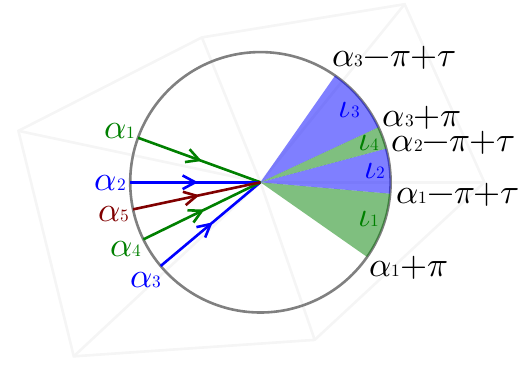}
	\caption{Incoming angles $\alpha_i$ and outgoing angle ranges $\iota_i$. $\iota_5 = \emptyset$}
	\label{vertex_figure_5}
\end{figure}

Let us explain how it works in the example shown in Figure~\ref{vertex_figure_5}. In this figure, the five geodesics $g_1, \dots, g_5$ of incoming angles $\alpha_1, \dots, \alpha_5$ (respectively) arrive at the vertex in this order. Here we assume $\alpha_1 < \alpha_2 < \alpha_5 < \alpha_4 < \alpha_3 < \alpha_1 + \tau$. Each incoming angle $\alpha_i$ is mapped to the corresponding outgoing angle range $\iota_i = [\mu_i, \nu_i]$ as follows:

\begin{enumerate}
	\item $\alpha_1$: the outgoing angle range is $\iota_1 = [\alpha_1 + \pi, \alpha_1 - \pi + \tau]$ and corresponding one pseudo-source interval (not shown) is generated.
	\item $\alpha_2$: $\mu_2$ is limited by $g_1$ while $\nu_2$ is not, thus $\iota_2 = [\alpha_1 - \pi + \tau, \alpha_2 - \pi + \tau]$ and two corresponding pseudo-source intervals (not shown) are created.
	\item $\alpha_3$: neither $g_1$ nor $g_2$ limits $\iota_3$, thus $\iota_3 = [\alpha_3 + \pi, \alpha_3 - \pi + \tau]$.
	\item $\alpha_4$: $\iota_4$ is limited by $g_2$ and $g_3$, thus $\iota_4 = [\alpha_2 - \pi + \tau, \alpha_3 + \pi]$.
	\item $\alpha_5$: $\iota_5$ is limited by $g_2$ and $g_4$ thus $\iota_5$ is temporarily computed as $\iota_5 = [\alpha_2 - \pi + \tau, \alpha_4 + \pi]$. However, it is empty and no pseudo-source intervals are created.
\end{enumerate}

Since the outline, initialization and procedure for edge events (Algorithm~\ref{algo_main},~\ref{algo_init},~\ref{algo_edge_event}) are identical, we only explain the procedure for vertex events and geodesics query. All intervals generated in this way form a tree structure by the \emph{Parent} pointer. we call it the \emph{reduced geodesic interval tree} or the \emph{reduced GIT}.

\begin{remark}
	In the reduced geodesic interval tree, only pseudo-source intervals generated at a common vertex are ensured to be disjoint. There may be an overlap between two non-pseudo-source intervals, or between a pseudo-source interval and a non-pseudo-source interval.
\end{remark}

\subsection{Procedure for vertex events}
Like the naive version, when a vertex event occurs, the associated interval $I$ is recorded on $v$. If $v$ is hyperbolic, it performs the procedure of the reduction explained in the previous subsection, and, if the resulting outgoing angle range is not empty, pseudo-source intervals are generated and the associated edge events and vertex events (if exist) are inserted into the event queue. (Algorithm~\ref{algo_vertex_event_reduced})

\begin{algorithm}%[h]
	\caption{(Processing a vertex event, for the reduced GIT)}
	\label{algo_vertex_event_reduced}
	\begin{algorithmic}[1]
	\Function {VertexEvent.Handle} {$Q$}
		\State $I$ := the associated interval
		\State $v$ := the starting point of $I$
		\State Record $I$ on $v$
		\State \textbf{if} $v$ is not hyperbolic \textbf{then} \Return
		\State $\alpha$ := the incoming angle of the geodesic at $v$
		\State $\tau$ := the total angle of $v$
		\State $\delta := \tau - 2\pi$
		\If {there exist no incoming angles within $(\alpha - \delta, \alpha)$}
			\State $\mu := \alpha + \pi$
		\Else
			\State $\mu := (\mbox{the largest incoming angle within }(\alpha - \delta, \alpha)) - \pi + \tau$
		\EndIf
		\If {there exist no incoming angles within $(\alpha, \alpha + \delta)$}
			\State $\nu := \alpha - \pi + \tau$
		\Else
			\State $\nu := (\mbox{the smallest incoming angle within }(\alpha, \alpha + \delta)) + \pi$
		\EndIf
		\State \textbf{if} $\mu \ge \nu$ \textbf{then} \Return
		\State $I_1 \dots I_k :=$ the pseudo-source intervals for the outgoing angle range $[\mu, \nu]$
		\For {$i$ in $1 \dots k$}
			\State $Q$.\Call{Push}{EdgeEvent($I_i$)}
		\EndFor
		\For {$i$ in $2 \dots k$}
			\State $Q$.\Call{Push}{VertexEvent($I_i$)}
		\EndFor
	\EndFunction
	\end{algorithmic}
\end{algorithm}

\subsection{Geodesics enumeration query}
In this subsection, we discuss the geodesics enumeration query for the reduced geodesic interval tree, which inputs a point $t$ on $\mathcal P$ and outputs the set $G_{st}^R$ of directed geodesics from $s$ to $t$, whose length is less than $R$.

In the complete geodesic interval tree, a geodesic is given for each pair $(I, p)$ by the \textsc{ConstructGeodesic} function (Algorithm~\ref{algo_construct_geodesic}). On the other hand, in the reduced geodesic interval tree, geodesics of the same path between the last vertex $v$ and $p$ are given together for the pair $(I, p)$. The primitive geodesic between $vp$ is given by the \textsc{ConstructPrimitiveGeodesic} function (Algorithm~\ref{algo_primitive_geodesic}). The geodesic is constructed from $t$ to $s$, and every time it hits a hyperbolic vertex, possible branches must be enumerated.

\begin{figure}
	\centering
	\includegraphics[height=7cm, pagebox=cropbox]{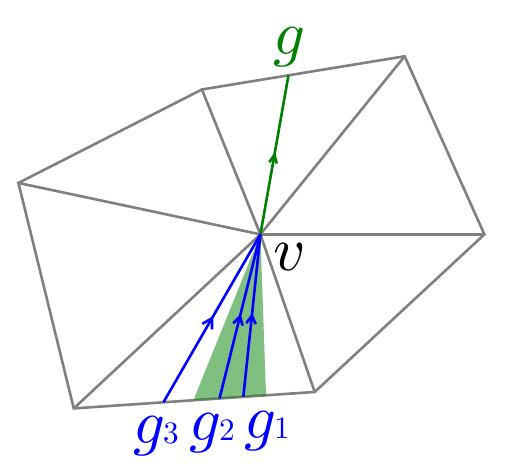}
	\caption{During the geodesics query, a geodesic is traced backwards. For the geodesic $g$ outgoing from $v$, we enumerate all connectable geodesics ($g_1$ and $g_2$) incoming into $v$ and check whether each can satisfy the upperbound. We recursively perform this process.}
	\label{vertex_figure_6}
\end{figure}

Let us take a look at the example illustrated in Figure \ref{vertex_figure_6}. In this figure, each arrow indicates the direction from $s$ to $t$, although the actual query is performed backwards. Let us assume that we have just found a geodesic $g$ outgoing from $v$, and there are three geodesics $g_1$, $g_2$ and $g_3$ incoming into $v$, and each of $g_1$ and $g_2$ is connectable with $g$ as a geodesic while $g_3$ is not. Then, for each of $g_1$ and $g_2$, we check it can satisfy the length upperbound, and if so, we connect it with $g$ and perform this process recursively.

In general, we can use a simple depth-first search here. For each interval $I$, $I.\Depth$ is the minimum of the depths of these geodesics grouped together. Since $G_{st}^R$ contains at least one geodesic represented by the pair $(I, t)$ if and only if the minimum of their lengths is less than $R$, the procedure of the \textsc{GetIntervals} function (Definition~\ref{def_get_intervals}) is identical. The whole procedure is given by Algorithm~\ref{algo_geodesic_query_reduced}. In the pseudocode, \textsc{RemoveLast} is the operation of removing the last point from the sequence, and is used for endpoints of primitive geodesics not to appear twice.

\begin{algorithm}
	\caption{(Construction of a primitive geodesic)} \label{algo_primitive_geodesic}
	\begin{algorithmic}[1]
	\Function {ConstructPrimitiveGeodesic} {$I$, $p$} // $p \in I.\Face$
		\State $g$ := $(p)$
		\State // when reached $s$, $I.\Parent =$ Null
		\State // when reached a hyperbolic vertex, $I.\Parent.\Depth < I.\Depth$
		\While {$I.\Parent \ne$ Null and $I.\Parent.\Depth = I.\Depth$}
			\State $e$ := $I.\Parent.\Edge$
			\State $p$ := \Call{Intersect}{$e$, the line segment connecting $p$ and $I.\Center$}
			\State $g$.\Call{AddFront}{$p$}
			\State $I$ := $I.\Parent$
		\EndWhile
		\State $g$.\Call{AddFront}{$I.\Center$}
		\Comment $I.\Center$ is $s$ or a hyperbolic vertex
		\State \Return ($g$, $I.\Parent =$ Null)
		\Comment tuple of a path and a Boolean value
	\EndFunction
	\end{algorithmic}
\end{algorithm}

\begin{algorithm}%[h]
	\caption{(Geodesics enumeration query, for the reduced GIT)} \label{algo_geodesic_query_reduced}
	\begin{algorithmic}[1]
	\Function {GeodesicEnumQuery} {$t$}
		\State $G$ := $\emptyset$
		\For {$I$ in \Call{GetIntervals}{$t$}} \Comment Definition~\ref{def_get_intervals}
			\State $d$ := \Call{Distance}{$t$, $I.\Center$}
			\Comment length of the primitive geodesic
			\State \Call{GeodesicEnumQueryRec} {$G$, $(t)$, $I$, $t$, $d$}
		\EndFor
		\State \Return $G$
	\EndFunction
	\end{algorithmic}
	\begin{algorithmic}[1]
	\Function {GeodesicEnumQueryRec} {$G$, $g$, $I$, $p$, $d$}
		\State ($h$, \textsc{IsSource}) := \Call{ConstructPrimitiveGeodesic} {$I$, $p$}
		\Comment Algorithm~\ref{algo_primitive_geodesic}
		\State $h$.\Call{RemoveLast}{\null}
		\If {\textsc{IsSource}}
			% \State $g$.AddFront($s$)
			\State $G$.\Call{Add}{$h + g$}
			\Comment $+$ denotes concatenation of sequences
			\State \Return
		\EndIf
		\State $v$ := the starting point of $h$
		\Comment this is a hyperbolic vertex
		\State $\alpha$ := the outgoing angle of $h$ at $v$
		\ForAll {$J$ : the intervals of incoming angle within $[\alpha + \pi, \alpha - \pi + \tau]$ at $v$}
			\State $l$ := \Call{Distance}{$v$, $J.\Center$}
			\Comment length of the primitive geodesic
			\If {$d + l + J.\Depth < R$}
				\State \Call{GeodesicEnumQueryRec} {$G$, $h + g$, $J$, $v$, $d + l$}
				\Comment recursive call
			\EndIf
		\EndFor
	\EndFunction
	\end{algorithmic}
\end{algorithm}

\subsection{Constructing single-pair geodesic graph}
\begin{figure}%[h]
	\centering
	\includegraphics{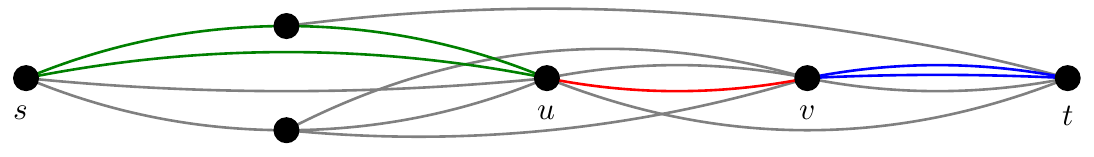}
	\caption{Concept of geodesic graph. Each arc represents a primitive geodesic. The red arc is connectable to the green arcs and the blue arcs.}
	\label{fig_geodesic_graph_concept}
\end{figure}

Since a geodesic is a sequence of primitive geodesics, we can reduce $G_{st}^R$ into a graph whose edges are primitive geodesics (Figure~\ref{fig_geodesic_graph_concept}). We call it a \emph{single-pair geodesic graph}, or simply a \emph{geodesic graph}. It can be computed directly using the reduced geodesic interval tree.

\begin{definition}[Single-pair geodesic graph]
	The \emph{(single-pair) geodesic graph} $\mathcal G_{st}^R$ with respect to $G_{st}^R$, is the directed graph satisfying the following conditions:
	\begin{itemize}
		\item The vertices of $\mathcal G_{st}^R$ are $s, t$ and the vertices of $\mathcal P$ through which at least one geodesic in $G_{st}^R$ passes.
		\item The edges of $\mathcal G_{st}^R$ are the directed primitive geodesics connecting two vertices of $\mathcal G_{st}^R$, such that each of them is contained in at least one directed geodesic in $G_{st}^R$.
	\end{itemize}
\end{definition}

\begin{remark}
	\label{rem_source_sink}
	In a geodesic graph $\mathcal G_{st}^R$, $s$ and $t$ act as the source and the sink (respectively). Even if $s$ (resp. $t$) is placed at a vertex and geodesics pass through the vertex, we regard $s$ (resp. $t$), as a vertex of $\mathcal G_{st}^R$, to be distinct from any other vertex of $\mathcal P$. This convention allows us to design the algorithm without special treatment of $s$ (resp. $t$) being a vertex or not.
\end{remark}

\begin{remark}
	A geodesic graph can have multi-edges. That is, there may exist pairs of edges such that the both endpoints coincide.
\end{remark}

\begin{remark}
	Our geodesic graph bears some resemblance to the Saddle Vertex Graph (SVG)~\cite{SVG}, but fundamentally differs as follows:
	\begin{itemize}
		\item The SVG only considers shortest geodesics, while our geodesic graph considers non-shortest geodesics as well.
		\item The SVG considers shortest geodesics for all pairs of vertices, while our geodesic graph only considers geodesics connecting $s$ and $t$.
		\item More importantly, the SVG is constructed before their geodesic query and the query is performed on the SVG, while our geodesic query is performed on the reduced geodesic interval tree and yields a geodesic graph. In other words, our geodesic graph is a representation of the query result.
	\end{itemize}
\end{remark}

\begin{remark}
	Our geodesic graph $\mathcal G_{st}^R$ is an exact representation of the query result for the given $s$, $t$ and $R$. The naive representation $G_{st}^R$ often have many overlapped sub-geodesics. In this situation, we can compress them into a graph, eliminating redundancy that can be combinatorially reconstructed from the graph.
\end{remark}

\begin{remark}
	Every figure in this paper that renders computed geodesics (such as Figure \ref{torus} and Figure \ref{pumpkin}) is a 3D-rendered version of a geodesic graph presented in this subsection.
\end{remark}

A geodesic graph is constructed from $t$ to $s$ using Algorithm~\ref{algo_geodesic_graph}. We use a modified version of Dijkstra's algorithm to determine the shortest possible length from $t$ for each primitive geodesic. We have two points to remark:
\begin{itemize}
	\item When $t$ is given, an interval can yield at most two primitive geodesics, and each primitive geodesic can be specified by the pair $(I, \mbox{\sc IsTarget})$ where $I$ is an interval and \textsc{IsTarget} is a Boolean value indicating whether the primitive geodesic ends with $t$ (otherwise it ends with the starting point of $I$). Note that \textsc{IsTarget} is set to be \textsc{True} only through the initialization of the algorithm since we regard $t$ to be a vertex of $\mathcal G_{st}^R$ distinct from any other vertex (Remark~\ref{rem_source_sink}). Since we can assume that no duplicated intervals are supplied by the \textsc{GetIntervals} function, we need to check visitedness only when \mbox{\sc IsTarget} is set to \textsc{False} (the line 11--14).
	\item In the line 23, $l$ is the length of the primitive geodesic given by $(J, \mbox{\sc False})$. A conceptual figure is described in Figure~\ref{fig_geodesic_graph_concept}. Let the red curve be the primitive geodesic and denoted as $h$. Not necessarily all geodesics in $G_{st}^R$ contain $h$. Let the green part and blue part be the subgraph of $G_{st}^R$ between $su$ and $vt$ respectively through which a geodesic containing $h$ can pass (in general the green and blue subgraph may be overlapped). Then, the distance between $su$ on the green subgraph is $J.\Depth$, because of the construction method of the reduced geodesic interval tree. Moreover, the distance between $vt$ on the blue subgraph is $d$, because of the construction algorithm of the geodesic graph, which is derived from Dijkstra's algorithm. Therefore the minimum length of geodesics between $st$ containing $h$ is $J.\Depth + l + d$, and $h$ is contained in $\mathcal G$ as an edge if and only if it is less than $R$.
\end{itemize}

%Aside from the event queue, a priority queue $Q_r$ is used to construct a geodesic graph, and processes in the ascending order of the length from $t$. A geodesic graph can be computed by Algorithm~\ref{algo_geodesic_graph}. Its outline is:

% \begin{enumerate}
% 	\item Let $Q_r$ be a priority queue, $\mathcal I_{\rm vis} := \emptyset$ to store visited intervals and $\mathcal G := \emptyset$ to store the geodesic graph.
% 	\item Insert the intervals of the geodesics reaching $t$ into $Q_r$;
% 	\item Take the interval $I$ of the geodesic of the minimum length from $t$ out of $Q_r$;
% 	\item If this instance of $I$ is inserted at (6), visited checking is performed as follows: If $\mathcal I_{\rm vis}$ contains $I$, skip (5) and (6). otherwise, add $I$ to $\mathcal I_{\rm vis}$ (See Remark \ref{rem_visited_intervals});
% 	\item Obtain the primitive geodesic using Algorithm~\ref{algo_primitive_geodesic};
% 	\item Add the primitive geodesic obtained in (4) to $\mathcal G$. If it did not reach the root of the tree, insert the unvisited intervals of the geodesics with connectable incident angle and length satisfying the upperbound into $Q_r$;
% 	\item Return to (2) until $Q_r$ becomes empty.
% \end{enumerate}

\begin{algorithm}%[h]
	\caption{(Construction of a geodesic graph, for the reduced GIT)} \label{algo_geodesic_graph}
	\begin{algorithmic}[1]
	\Function {ConstructGeodesicGraph} {$t$}
		\State $Q_r$ := a priority queue
		\State $\mathcal I_{\rm vis}$ := $\emptyset$ (the set of visited intervals)
		\State $\mathcal G$ := $\emptyset$ (the set of edges of the output geodesic graph)
		\For {$I$ in \Call{GetIntervals}{$t$}} \Comment Definition~\ref{def_get_intervals}
			\State $d$ := \Call {Distance}{$t$, $I.\Center$} \Comment length of the primitive geodesic
			\State $Q_r$.\Call{Push}{$(I, \mbox{\sc True}, \mbox{priority: } d)$}
		\EndFor
		\While {$Q_r$ is not empty}
			\State $(I, \mbox{\sc IsTarget}, \mbox{priority: } d)$ := $Q_r$.Pop()
			\If {\textbf{not} \textsc{IsTarget}}
			% \Comment Remark~\ref{expl_visited_intervals}
				\State \textbf{if} $\mathcal I_{\rm vis}$ contains $I$ \textbf{then continue}
				\State $\mathcal I$.\Call{Add}{$I$}
			\EndIf
			% \If {IsTarget}
			% 	\State $p$ := $t$
			% \Else
			% 	\State $p$ := the starting point of $I$
			% \EndIf
			\State $p$ := \textbf{if} \textsc{IsTarget} \textbf{then} $t$ \textbf{else} the starting point of $I$
			\State ($g$, \textsc{IsSource}) := \Call{ConstructPrimitiveGeodesic}{$I$, $p$}
			\Comment Algorithm~\ref{algo_primitive_geodesic}
			\State $\mathcal G$.\Call{AddEdge}{$g$}
			% \State \textbf{if} the starting point of $h$ is $s$ \textbf{then continue}
			\State \textbf{if} \textsc{IsSource} \textbf{then continue}
			% \If {the starting point of $h$ is a hyperbolic vertex $v$}
			% \If {not IsSource}
				\State $v$ := the starting point of $h$
				\Comment this is a hyperbolic vertex
				\State $\alpha$ := the outgoing angle of $g$ at $v$
				\ForAll {$J$ : the intervals of incoming angle within $[\alpha + \pi, \alpha - \pi + \tau]$ at $v$}
					\State $l$ := \Call{Distance}{$v, J.\Center$}
					\If {($\mathcal I_{\rm vis}$ does not contain $J$) and ($d + l + J.\Depth < R$)}
					% \Comment Remark~\ref{expl_distance_checking}
						\State $Q_r$.\Call{Push}{$(J, \mbox{False}, \mbox{priority: } d + l)$}
					\EndIf
				\EndFor
			% \EndIf
		\EndWhile
		\State \Return {$\mathcal G$}
	\EndFunction
	\end{algorithmic}
\end{algorithm}

\section{Performance}
In this section, since we evaluate both the naive version that generates a complete geodesic interval tree and the improved version that generates a reduced geodesic interval tree, we call them \emph{geodesic interval trees}.
% The former consumes much time, and much memory to retain the tree, while the latter only requires negligible amount of time and memory compared with former, thus we mainly evaluate the former.
Since the size of an output geodesic interval tree greatly depends on the geometry, it is difficult to express it only in terms of input size and $R$. The efficiency of a reduced geodesic interval tree is due to its small size compared with the corresponding complete one, and we evaluate it by experiments in the next subsection. First, we state an output-sensitive complexity evaluation:

\begin{theorem}
Let $\mathcal T$ be a (complete or reduced) geodesic interval tree and $N = |\mathcal T|$ be the number of intervals in it. The corresponding algorithm for generating $\mathcal T$ runs in $O(N \log N)$ time and $O(N)$ space.
\begin{proof}
Since the number of the whole generated intervals is $N$, the numbers of edge events and vertex events are $O(N)$. Thus, the size of the event queue is $O(N)$ at any time. Therefore, the pop operation of the event queue takes $O(\log N)$ time per event. Concerning processing time of an event, an edge event can be processed in constant time. A vertex event requires time proportional to the number of the intervals it generates, but it sums up to $O(N)$. Also, in the reduced version, the two adjacent incoming angles of an incoming angle must be acquired, but it can be done in $O(\log N)$ time per event using a balanced binary search tree. Therefore, the time complexity is $O(N \log N)$. On the other hand, since the required space is the whole output tree $\mathcal T$ and the event queue, and each interval or event consumes constant space, the space complexity is $O(N)$.
\end{proof}
\end{theorem}

\subsection{Experimental Result}
\begin{figure}
	\begin{minipage}{0.49\textwidth}
		\centering
		\includegraphics[height=5cm]{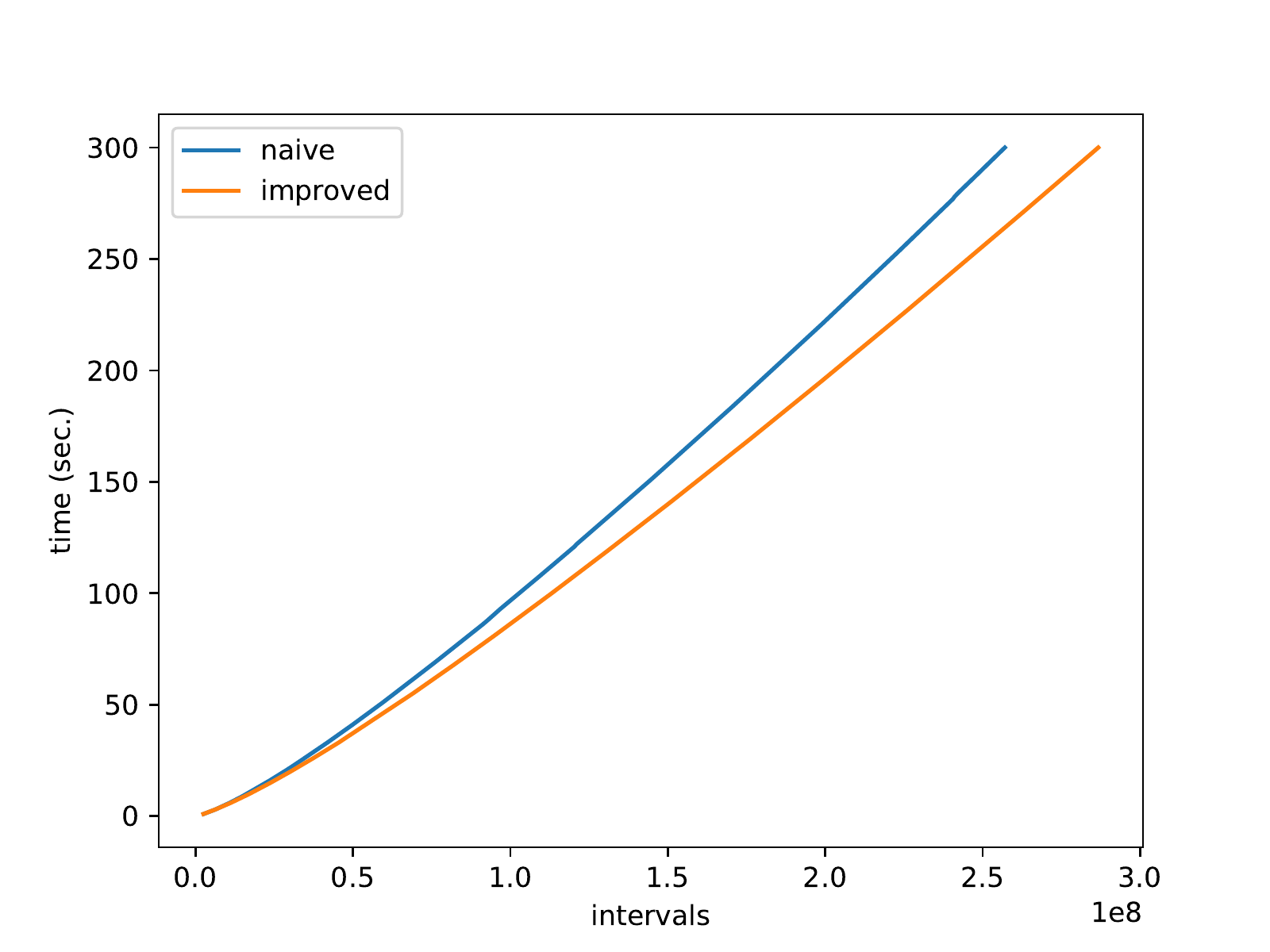}
		\caption{Time vs $|\mathcal T|$}
		\label{plot_time_intervals_elephant_0}
	\end{minipage}
	\begin{minipage}{0.49\textwidth}
		\centering
		\includegraphics[height=5cm]{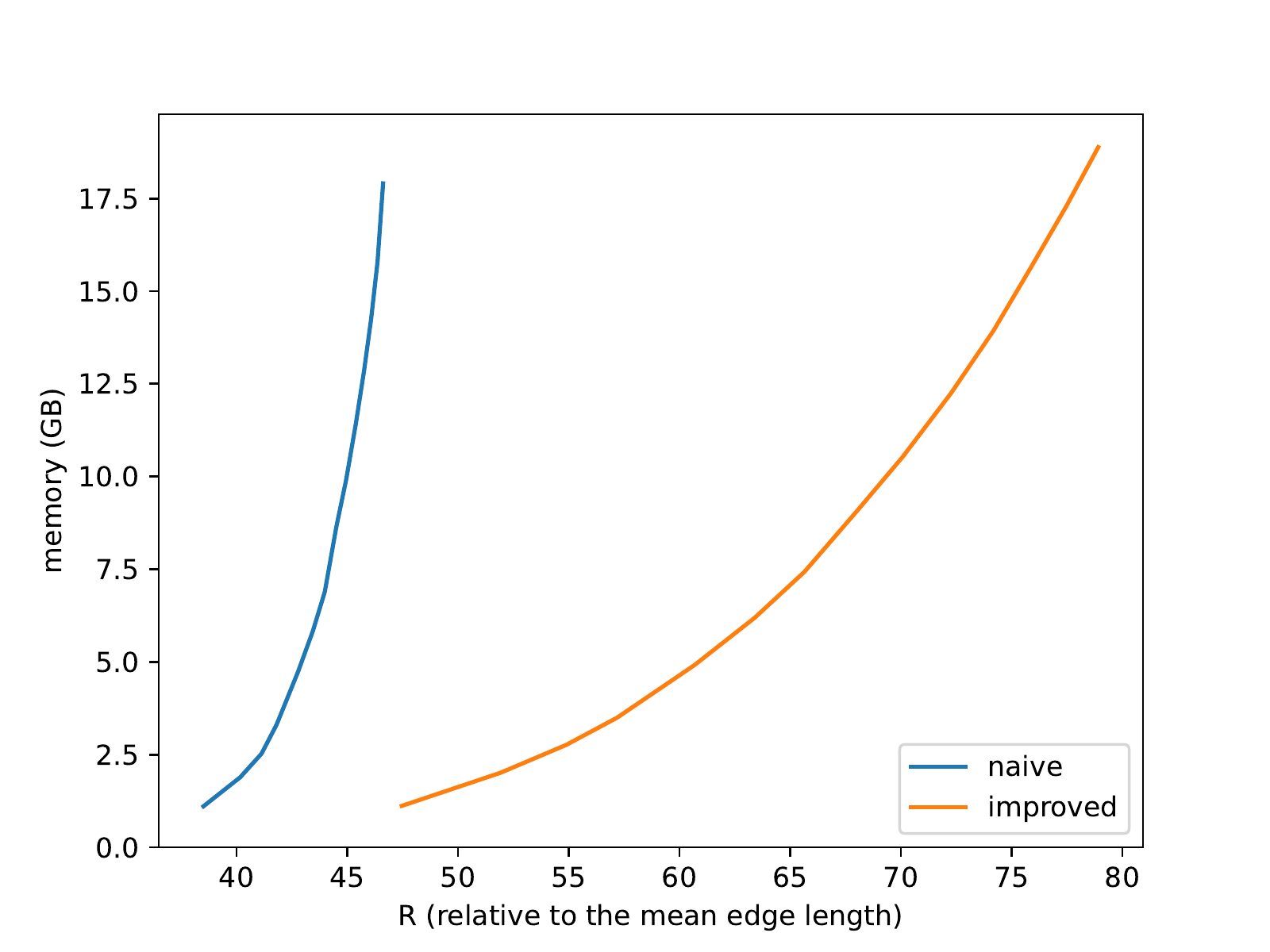}
		\caption{Memory vs $R$}
		\label{plot_memory_elephant_0}
	\end{minipage}
	\begin{minipage}{0.49\textwidth}
		\centering
		\includegraphics[height=5cm]{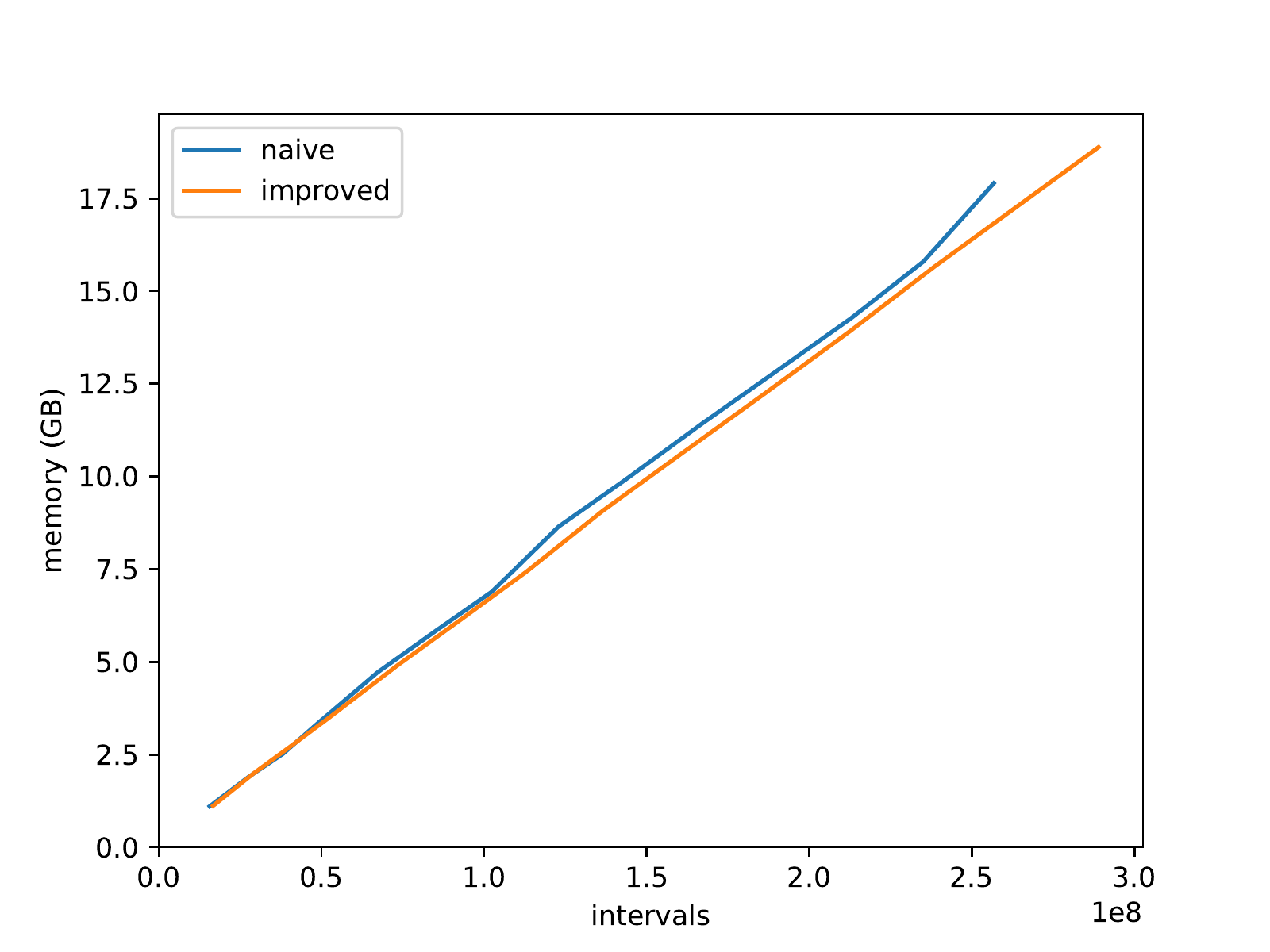}
		\caption{Memory vs $|\mathcal T|$}
		\label{plot_memory_intervals_elephant_0}
	\end{minipage}
	\begin{minipage}{0.49\textwidth}
		\centering
		\includegraphics[height=5cm]{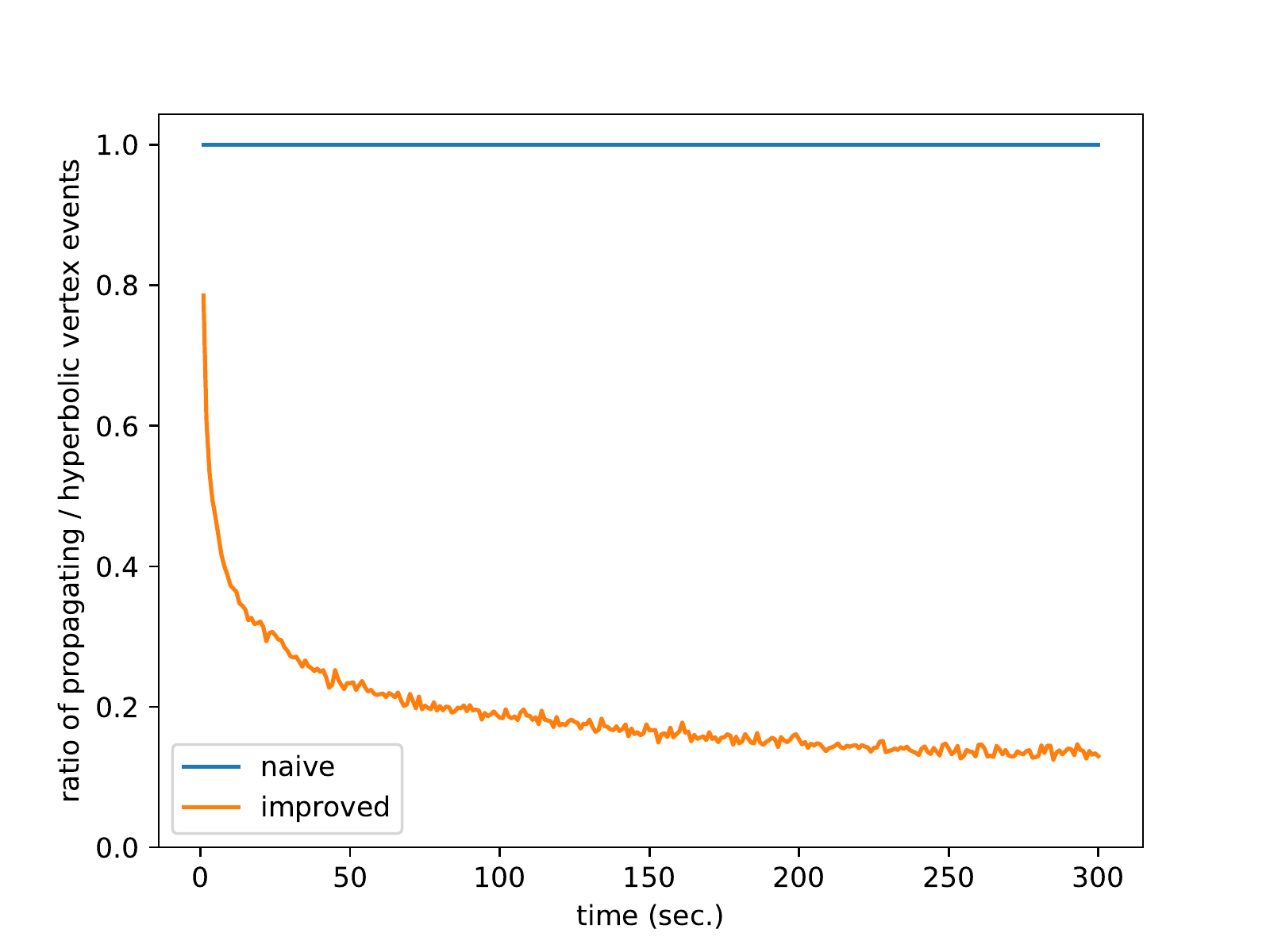}
		\caption{Ratio of the propagating vertex events to the hyperbolic ones among newly-processed events in one second}
		\label{plot_ratio_propagating_ve_elephant_0}
	\end{minipage}
	\begin{minipage}{0.49\textwidth}
		\centering
		\includegraphics[height=5cm]{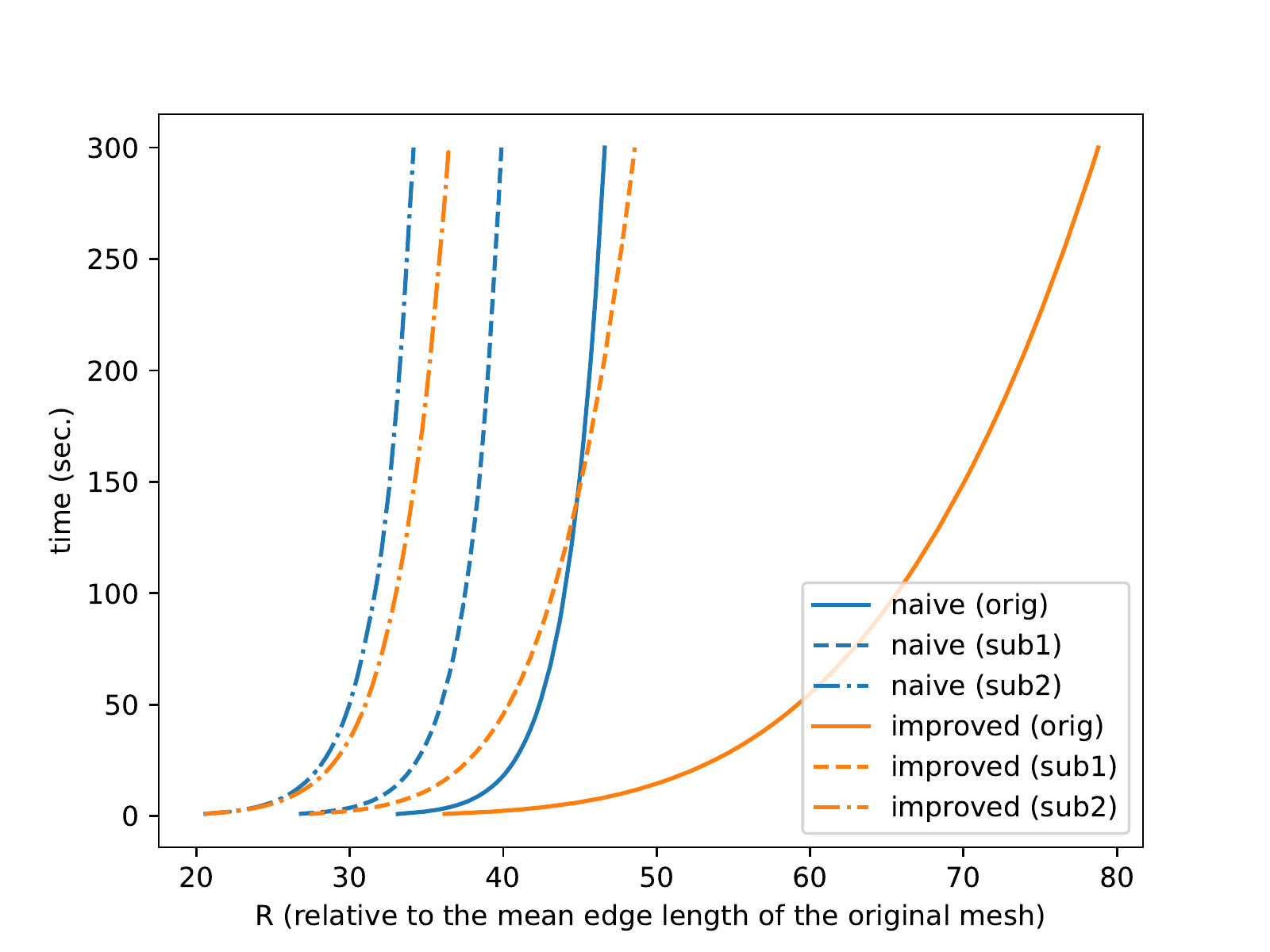}
		\caption{Subdivision and computation time of Elephant; sub1: 11K vertices, sub2: 44K vertices}
		\label{plot_subdivision_elephant}
	\end{minipage}
\end{figure}

\begin{figure}
	\centering %R = 0.696099, mem 3248020 KB (improved), 5721468 KB (naive)
	\includegraphics[height=8cm]{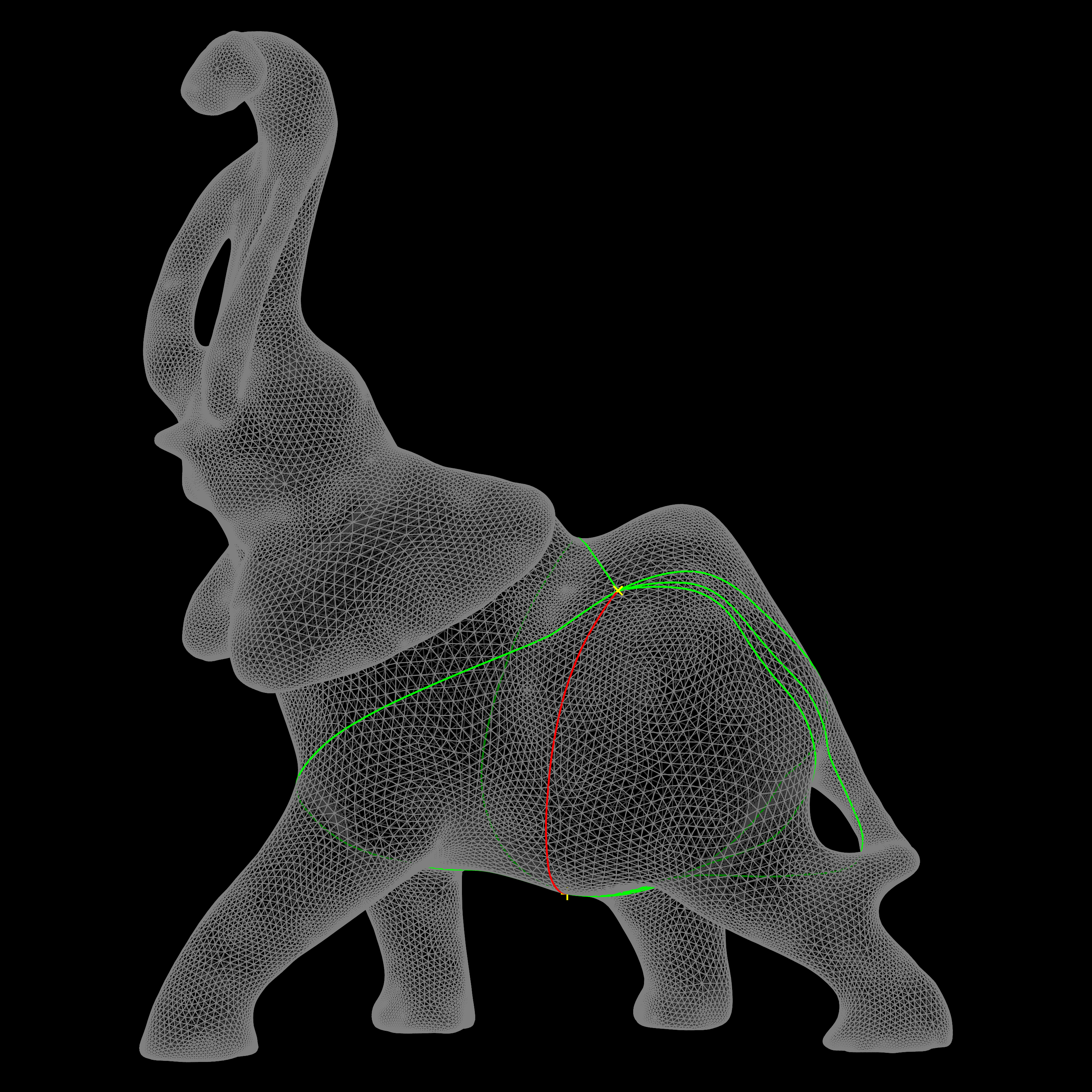}
	\caption{Surface (with 44460 vertices) obtained by subdividing twice from Elephant, $R = 131$ (relative to the mean edge length of this subdivided mesh); building the geodesic interval tree took 60.4 seconds (naive), 30.3 seconds (improved) and the total memory consumption (measured by the OS) is 5.5GB (naive), 3.1GB (improved); 83K intervals (naive), 46K intervals (improved) are generated; geodesics query took less than 1ms (both version).}
	\label{fig_elephant2_result}
\end{figure}

For the purpose of evaluating performance of our methods, we mainly use the Elephant mesh contained in the CGAL (Computational Geometry Algorithms Library)~\cite{CGAL}. The proposed methods consist of two parts, i.e., the construction of a geodesic interval tree, and the geodesics query or the construction of a geodesic graph. Since the geodesics query consumes much less (often negligible) time, we only evaluate the performance of the construction of a geodesic interval tree, except Figure~\ref{fig_elephant2_result}. We implemented our (single-threaded) algorithms in C++ and tested them using a machine with Intel(R) Core(TM) i9-9980XE CPU @ 3.00GHz and 128GB RAM running Linux. Except Figure~\ref{plot_memory_elephant_0},~\ref{plot_memory_intervals_elephant_0} and~\ref{fig_elephant2_result}, we ran our program for 300 seconds and took statistics every second. In Figure~\ref{plot_memory_elephant_0} and~\ref{plot_memory_intervals_elephant_0}, we manually recorded the amount of memory consumption measured by the OS, when it elapsed 10, 20, 30, 40, 60, 80, 100, 125, 150, 180, 210, 240, 270 and 300 seconds since each program started.

\begin{figure}
	\begin{minipage}{0.49\textwidth}
		\centering
		\includegraphics[height=5cm]{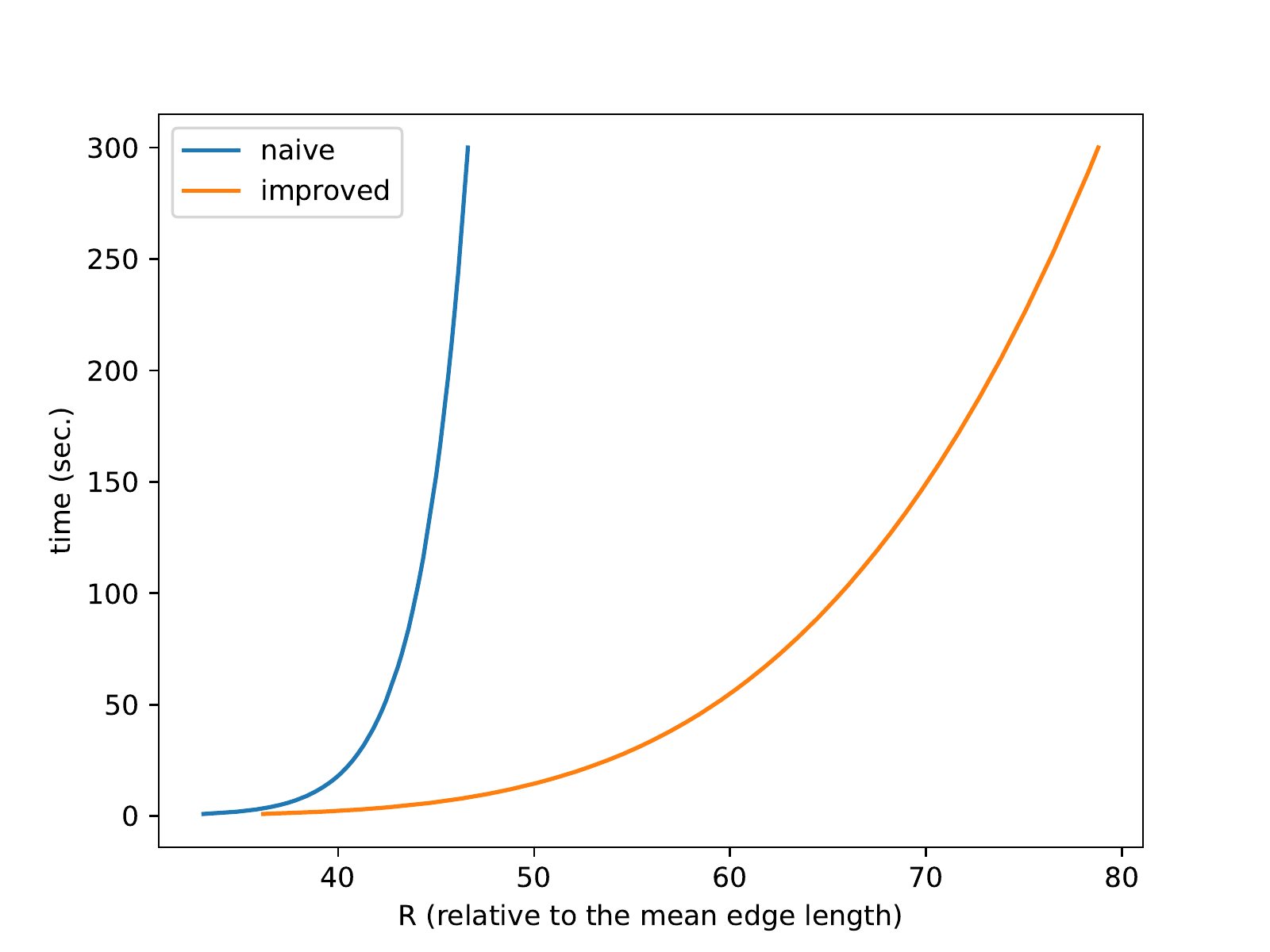}
		\caption{Linear plot of time vs $R$}
		\label{plot_time_elephant_0}
	\end{minipage}
	\begin{minipage}{0.49\textwidth}
		\centering
		\includegraphics[height=5cm]{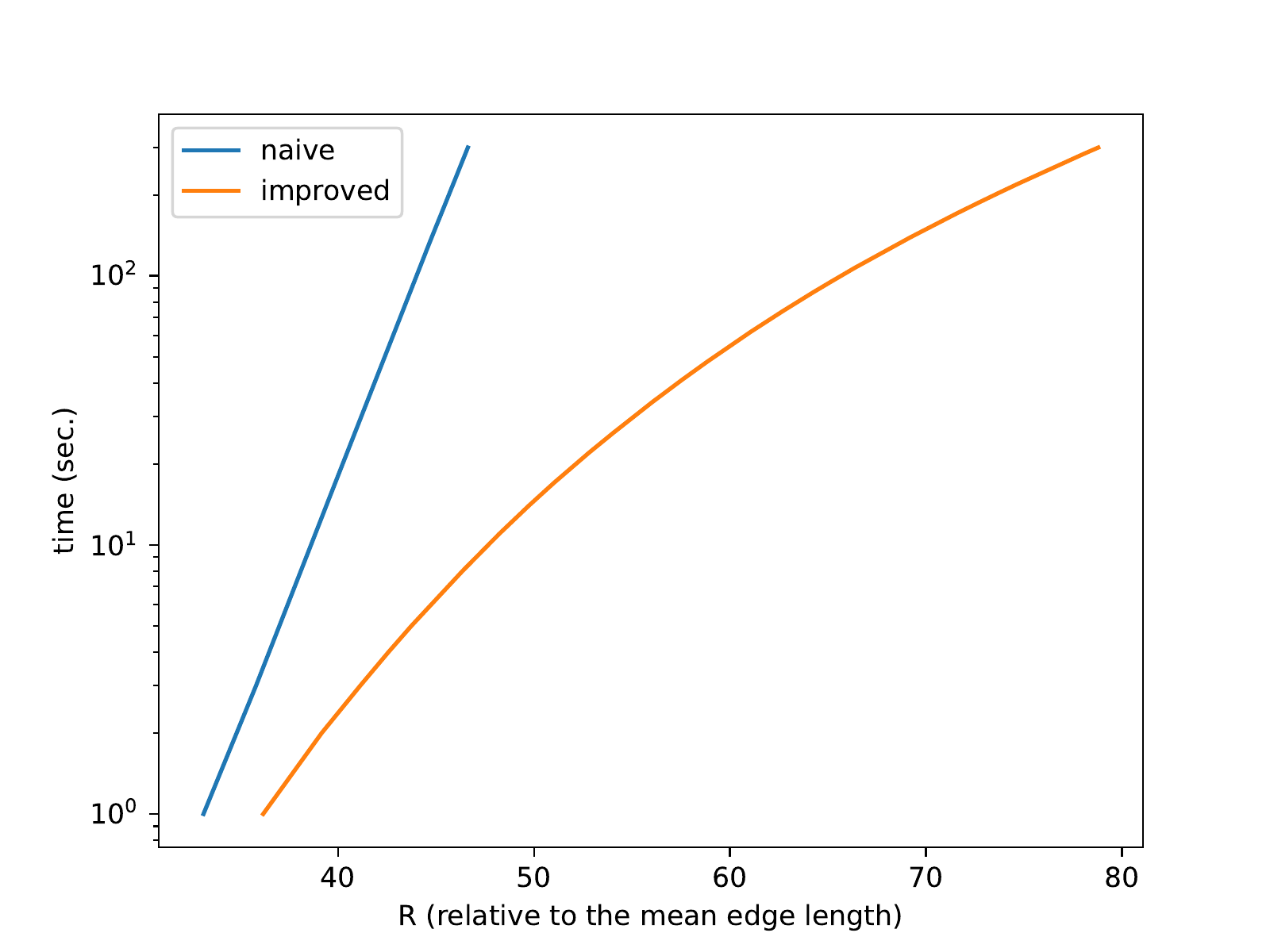}
		\caption{Log-linear plot of time vs $R$}
		\label{plot_log_time_elephant_0}
	\end{minipage}
	\begin{minipage}{0.49\textwidth}
		\centering
		\includegraphics[height=5cm]{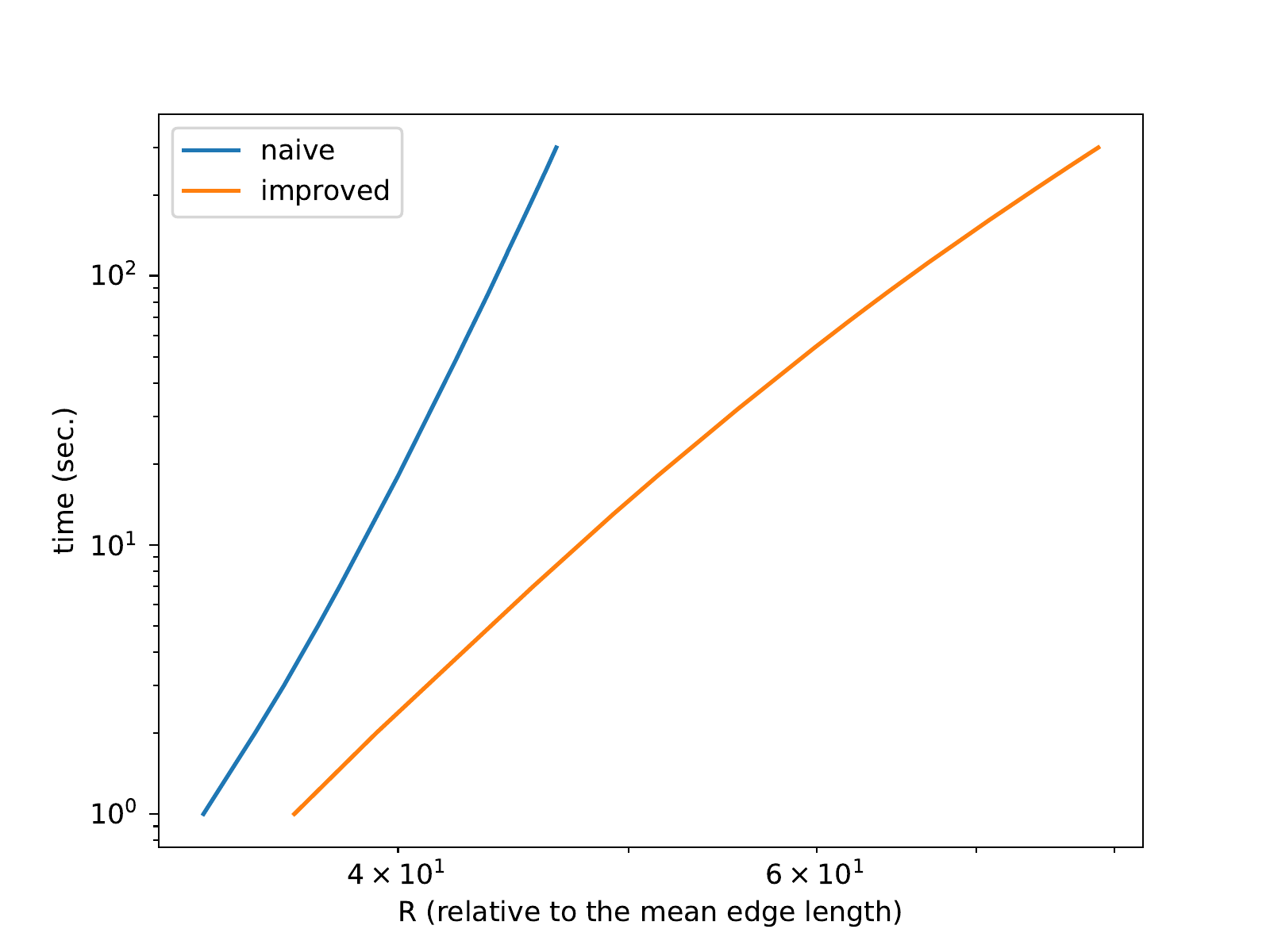}
		\caption{Log-log plot of time vs $R$}
		\label{plot_loglog_time_elephant_0}
	\end{minipage}
	\begin{minipage}{0.49\textwidth}
		\centering
		\includegraphics[height=5cm]{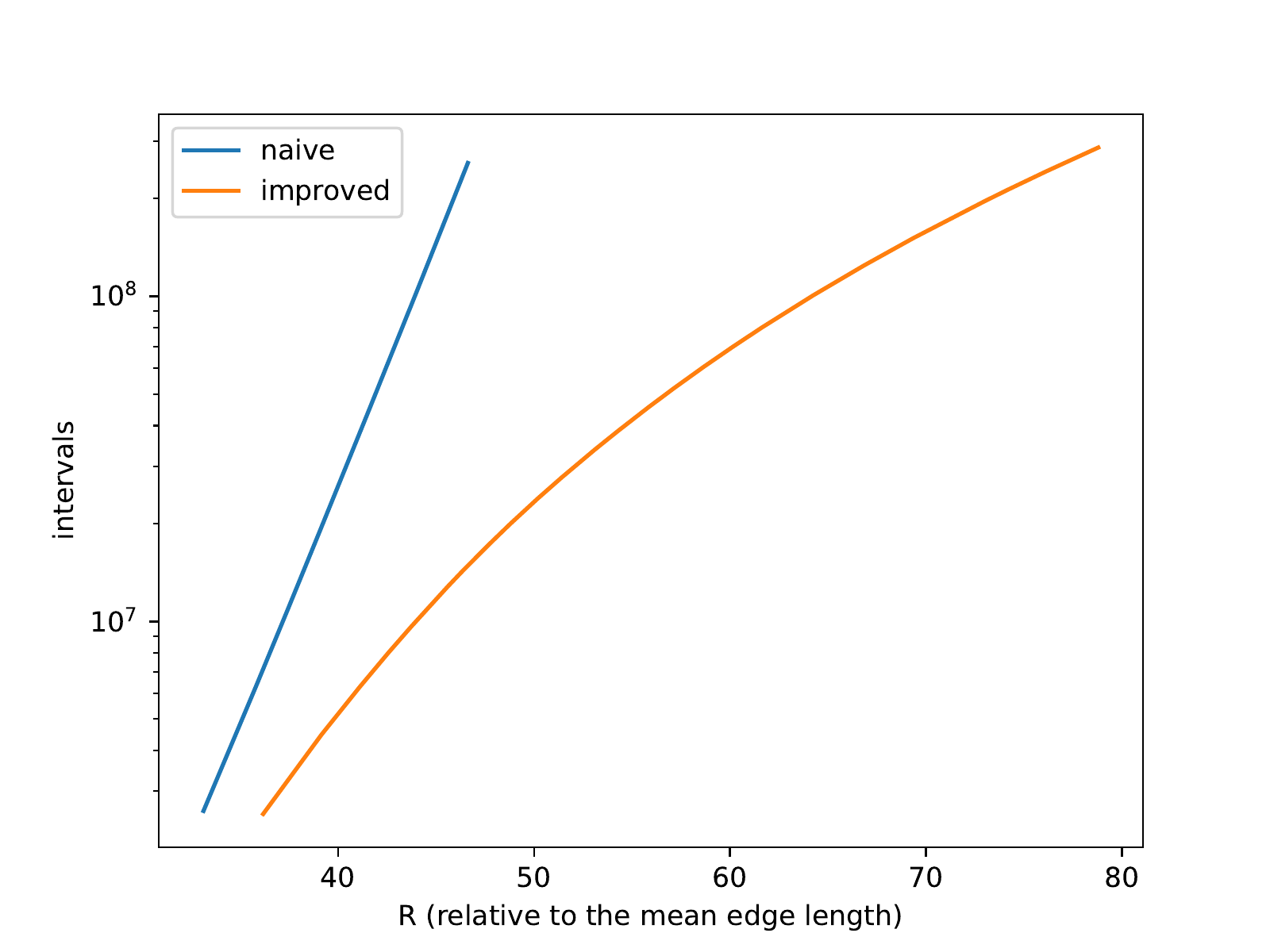}
		\caption{Log-linear plot of $|\mathcal T|$ vs $R$}
		\label{plot_log_intervals_elephant_0}
	\end{minipage}
	\begin{minipage}{0.49\textwidth}
		\centering
		\includegraphics[height=5cm]{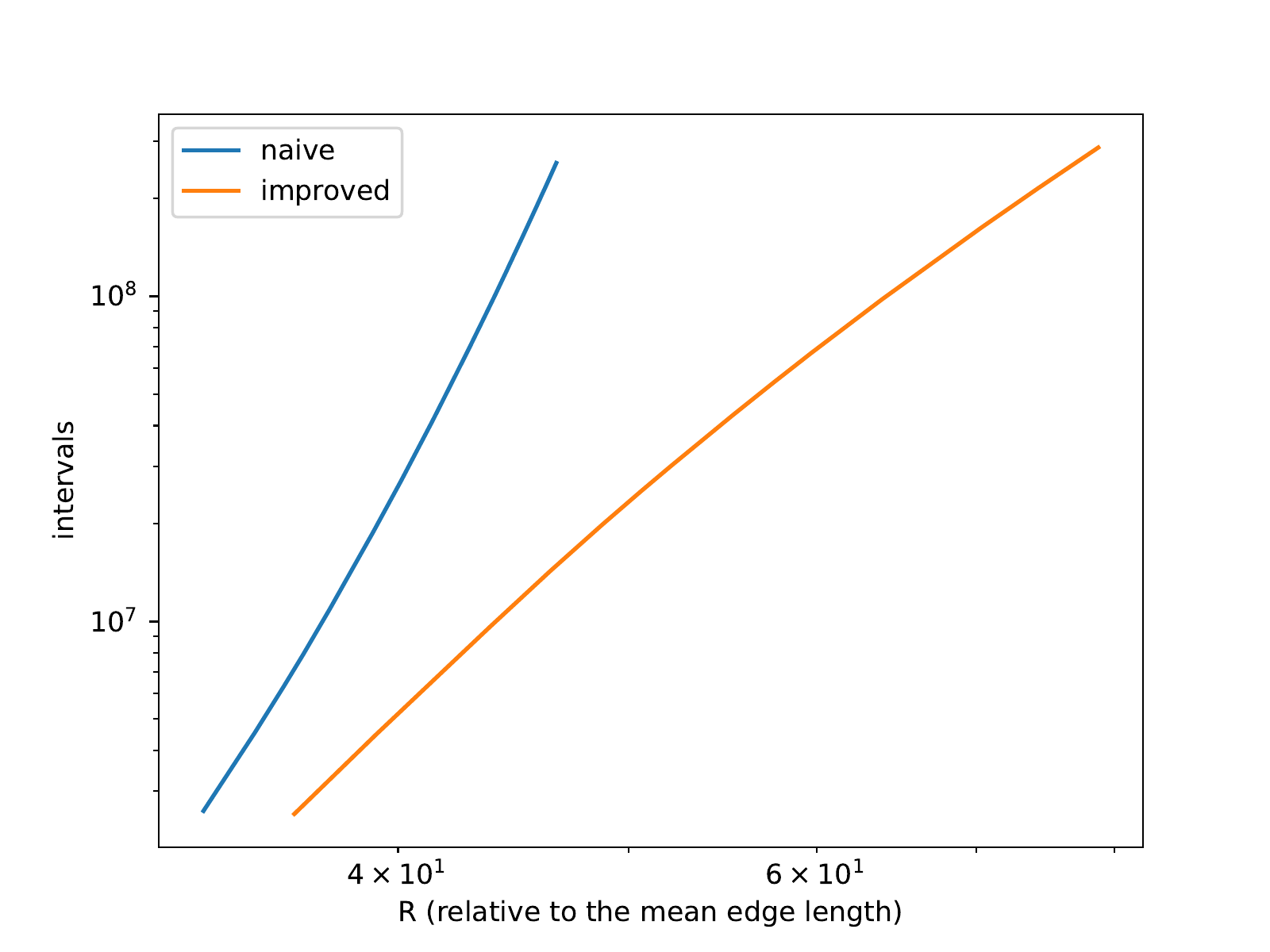}
		\caption{Log-log plot of $|\mathcal T|$ vs $R$}
		\label{plot_loglog_intervals_elephant_0}
	\end{minipage}
	\begin{minipage}{0.49\textwidth}
		\centering
		\includegraphics[height=5cm]{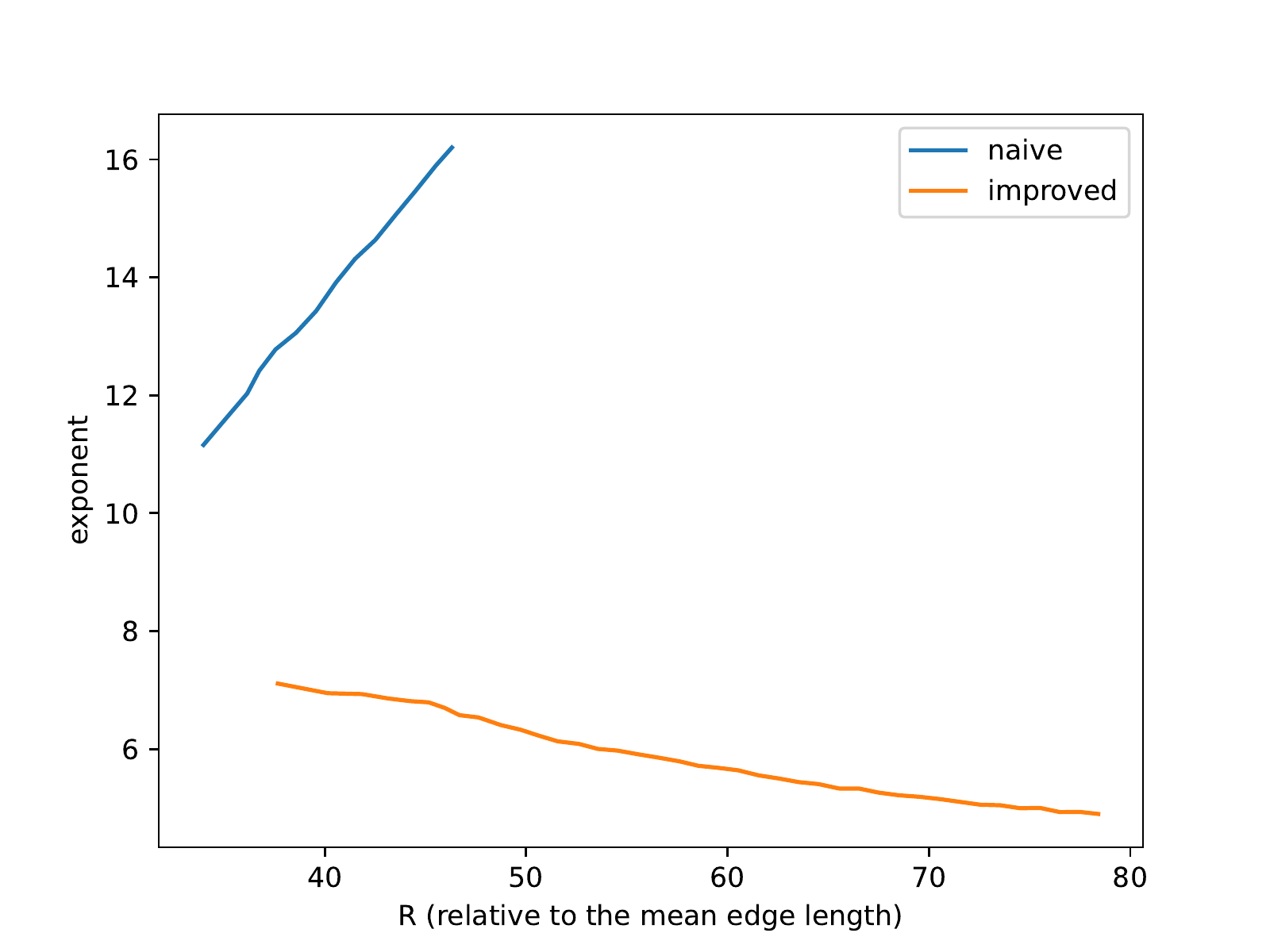}
		\caption{$\Delta(\log |\mathcal T|)/\Delta(\log R)$ vs $R$}
		\label{plot_diff_loglog_intervals_elephant_0}
	\end{minipage}
\end{figure}

Relation of $R$ (normalized so that the mean edge length is 1) and running time is shown in Figure~\ref{plot_time_elephant_0}, and its log-linear plot and log-log plot are shown in Figure~\ref{plot_log_time_elephant_0} and Figure~\ref{plot_loglog_time_elephant_0}. We can observe that the running time of the naive version grows exponentially to $R$, while that of improved version grows more slowly. A log-linear plot and log-log plot of the total number of generated intervals $|\mathcal T|$ are shown in Figure~\ref{plot_log_intervals_elephant_0} and Figure~\ref{plot_loglog_intervals_elephant_0}, and they exhibit a pattern similar to that of the computation time. Figure~\ref{plot_diff_loglog_intervals_elephant_0} shows the slope of the log-log plot, i.e. $\Delta(\log |\mathcal T|)/\Delta(\log R)$, which can be considered as the exponent $\alpha$ when $|\mathcal T|$ is locally fit by $kR^\alpha$ ($k$ and $\alpha$ are constants). This value is increasing in the naive version but decreasing in the improved version, as $R$ increases. Figure~\ref{plot_time_intervals_elephant_0} shows relation of $|\mathcal T|$ and computation time, and exhibits a pattern similar to the quasilinear growth. Memory consumption is shown in Figure~\ref{plot_memory_elephant_0} (to $R$) and Figure~\ref{plot_memory_intervals_elephant_0} (to $|\mathcal T|$). We can observe that the memory consumption seems to grow linearly to $|\mathcal T|$. Figure~\ref{plot_ratio_propagating_ve_elephant_0} shows the ratio of the number of propagating vertex events (vertex events that generated at least one interval) to the number of hyperbolic vertex events (vertex events that occurs at a hyperbolic vertex), among newly-processed events in one second. This value is 1 in the naive version but around 0.2 in the improved version in this instance, which means that, in the improved version, there are less vertex events that actually generate intervals compared with the naive version.

\begin{figure}
	\centering
	\includegraphics[height=5cm]{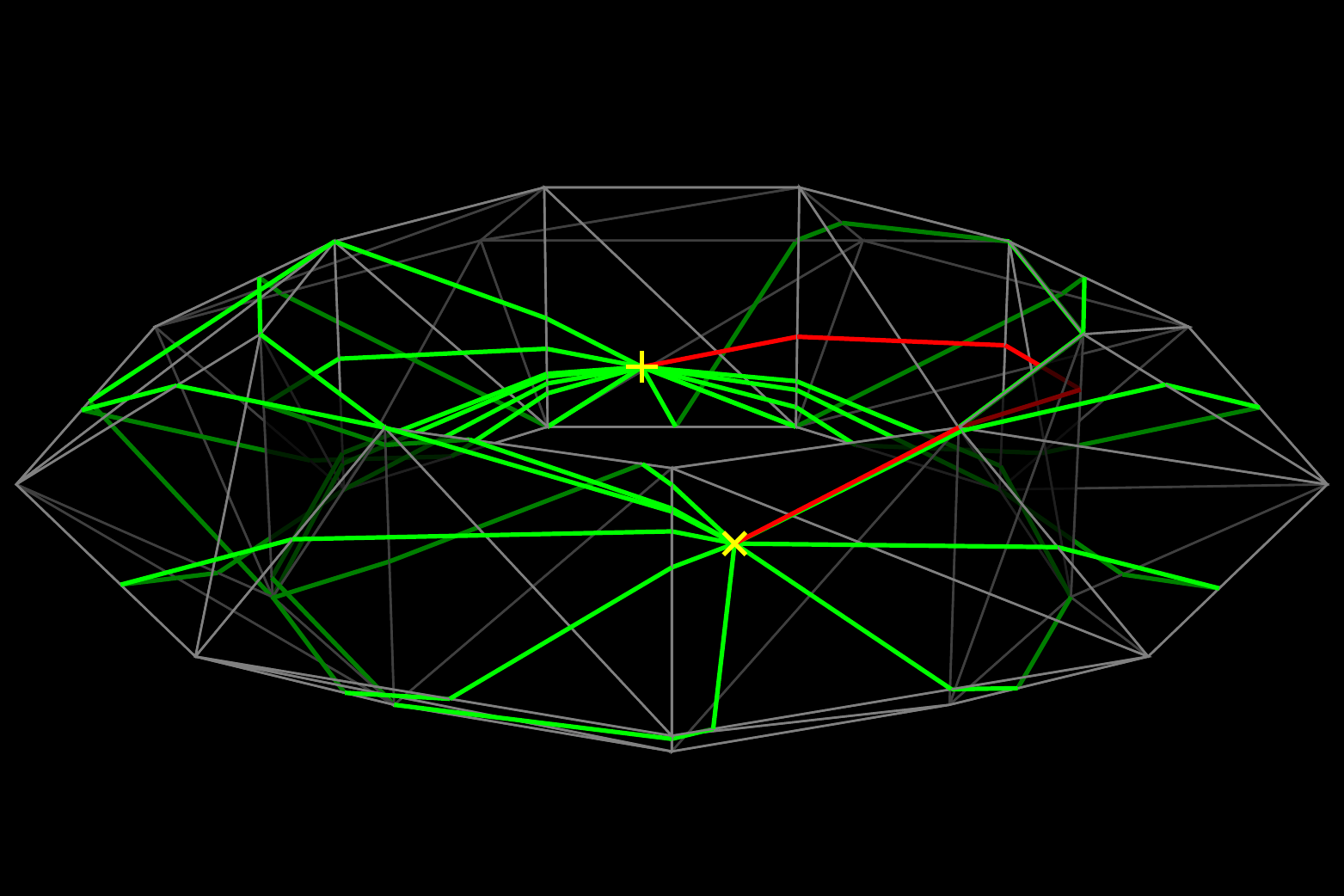}
	\caption{Simple Torus. Figures~\ref{plot_diff_loglog_intervals_naive_torus_3}--\ref{plot_diff_loglog_intervals_reduced_torus_3} are concerned on this mesh.}
	\label{fig_simple_torus}
\end{figure}

\begin{figure}
	\begin{minipage}{0.48\textwidth}
		\centering
		\includegraphics[height=5cm]{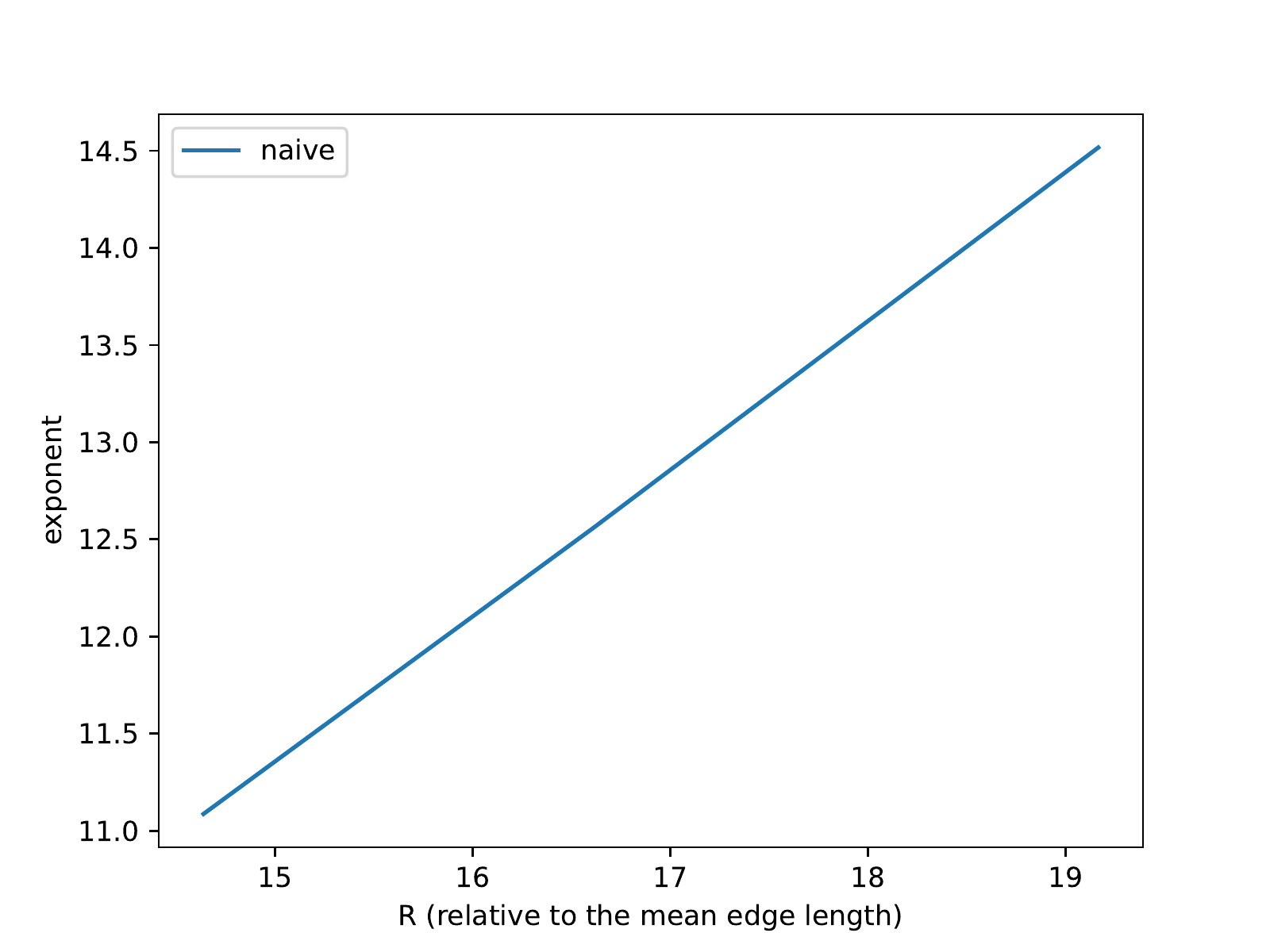}
		\caption{$\Delta(\log |\mathcal T|)/\Delta(\log R)$ vs $R$ in the naive version}
		\label{plot_diff_loglog_intervals_naive_torus_3}
	\end{minipage}
	\begin{minipage}{0.48\textwidth}
		\centering
		\includegraphics[height=5cm]{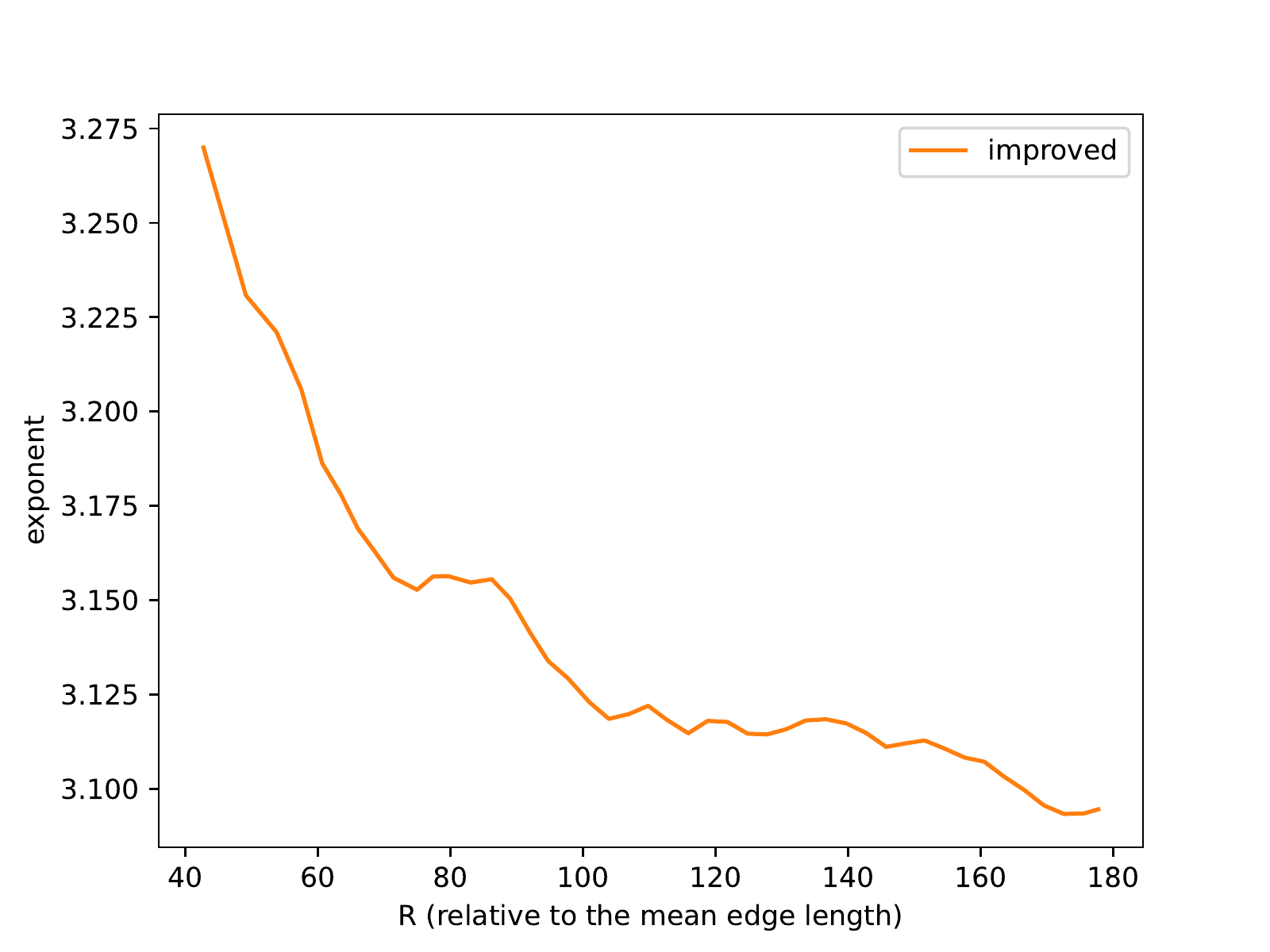}
		\caption{$\Delta(\log |\mathcal T|)/\Delta(\log R)$ vs $R$ in the improved version}
		\label{plot_diff_loglog_intervals_reduced_torus_3}
	\end{minipage}
\end{figure}

We also tested on a synthesized simple torus (Figure~\ref{fig_simple_torus},~\ref{plot_diff_loglog_intervals_naive_torus_3},~\ref{plot_diff_loglog_intervals_reduced_torus_3}). We can observe that, for sufficiently large $R$, the total number of generated intervals $|\mathcal T|$ in the improved version is approximately $\Theta(R^3)$. We have not established a theory, but we could \emph{guess} the reason behind it as follows:

\begin{itemize}
	\item For sufficiently (very) large $R$ (and in generic cases), pseudo-source intervals are likely to ``fill up'' all angles around hyperbolic vertices so that no new pseudo-source intervals are generated.
	\item In this situation, the area ``swept'' by the imaginary wavefront of intervals is $\Theta(R^2)$. That is, the number of vertex events is likely to be $\Theta(R^2)$. Since an interval in the wavefront splits into two intervals when it meets a vertex, the number of intervals in the wavefront is also likely to be $\Theta(R^2)$. Since $|\mathcal T|$ can be obtained by summing the size of each generation, it is likely to be $\Theta(R^3)$.
\end{itemize}

To evaluate relationship of smoothness of the mesh and performance, we used the recursively subdivided surfaces of Elephant using the Loop subdivision scheme~\cite{Loop}. In Figure~\ref{plot_subdivision_elephant}, the computation time is shown for the original mesh (orig), and the surface subdivided once (sub1) and twice (sub2). Here $R$ is relative to the mean edge length of the original mesh. We can see that the computation time of the improved version becomes closer to that of the naive version as the mesh becomes more detailed. The reason behind it is that, pseudo-source intervals become more unlikely to overlap in the naive version, since the total angle of each vertex becomes closer to $2\pi$. The surface subdivided twice (sub2) from Elephant is used in Figure~\ref{fig_elephant2_result} to evaluate practical performance. The detail of the experiment is explained in the caption of this figure, but compared with the naive version, we can observe that the improved version took nearly half computation time and used approximately 56\% memory to produce this result.

\begin{figure}
	\centering
	\includegraphics[height=8cm]{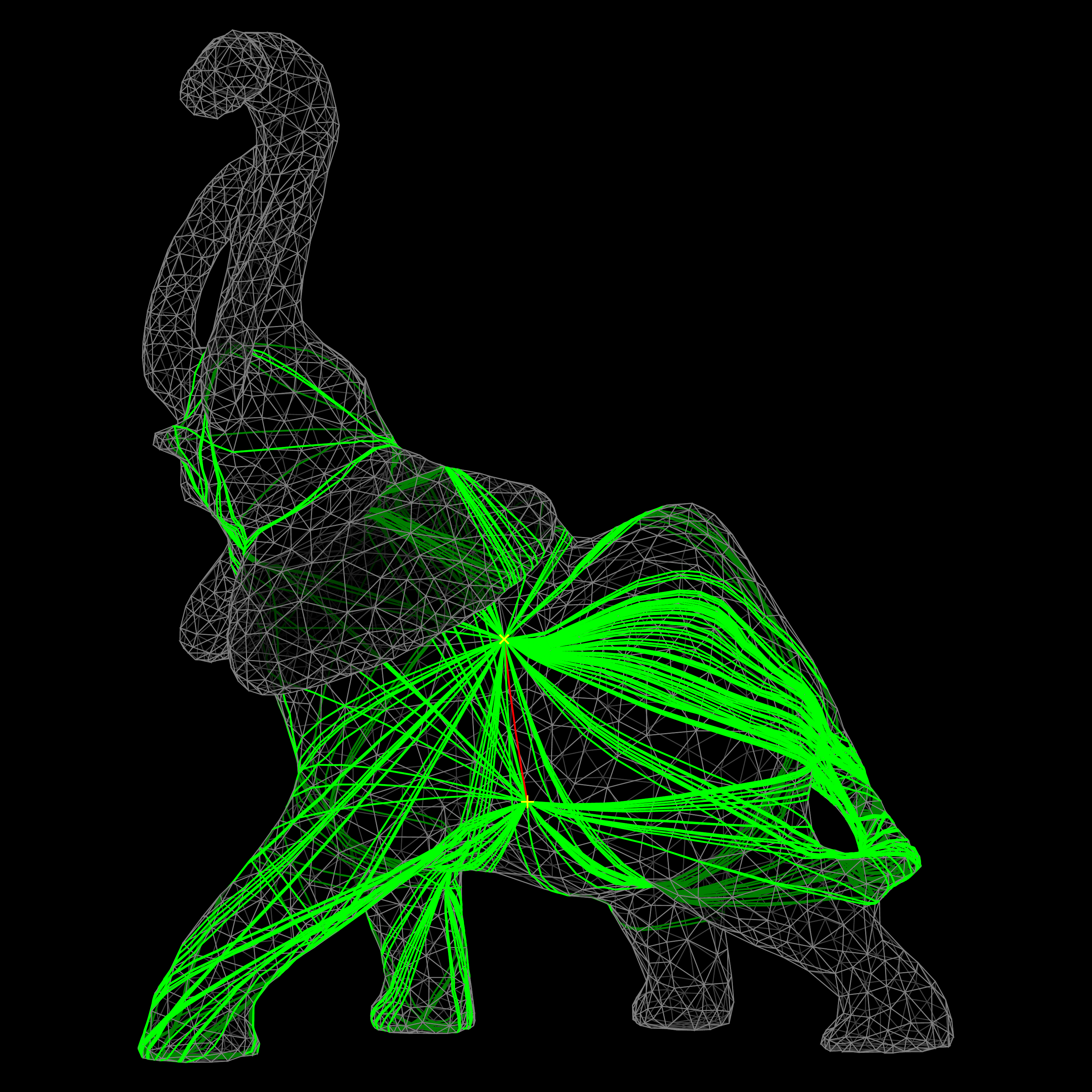}
	\caption{Elephant (2775 vertices and 5558 faces). Figures~\ref{plot_time_elephant_0}--\ref{plot_subdivision_elephant} are concerned on this mesh.}
	\label{fig_elephant_0}
\end{figure}

\section{Conclusion}
In this paper, we have formulated the single-source geodesics enumeration problem and presented two data structures for the problem: the complete geodesic interval tree and the reduced geodesic interval tree. Although we could not provide their worst-case bounds in terms of the input size, experiments suggested that the reduced geodesic interval tree is practically more efficient on a nonconvex polyhedron in terms of running time and memory consumption. On the assumption of reasonable complexity of the input mesh, our method allows geodesics to be computed in reasonable time, which opens up possibility of finding useful non-shortest geodesics which traditional shortest path algorithms cannot find.

We have given the complexity of our algorithms in terms of the size of the output geodesic interval trees. However, it largely depends on the geometry of the polyhedron, and a (non-trivial) complexity estimation is likely to include not only the input size (such as the number of vertices) and $R$, but also geometrical characteristics of the polyhedron (such as discretized curvatures). Therefore, rigorous analysis of complexity may involve mathematical study of geodesics themselves on a polyhedron. Geodesics yielded by our method can be used to approximate geodesics on a smooth surface, although more accurate or efficient algorithms tailored for smooth surfaces could be developed.

\section*{Acknowledgements}
The author sincerely appreciates Professor Hiroki Arimura and Professor Toru Ohmoto for giving many valuable advices.

%%%% bib %%%%%%%%%%%%%%%%%%%%%%%%%

\bibliographystyle{ws-ijcga}
\bibliography{ref}

\begin{thebibliography}{10}
\newcommand{\enquote}[1]{#1}

\bibitem{MMP}
J.~S.~B. Mitchell, D.~M. Mount and C.~H. Papadimitriou, \enquote{The discrete
  geodesic problem}, \emph{{SIAM} J. Comput.} \textbf{16} (1987) 647.

\bibitem{CH}
J.~Chen and Y.~Han, \enquote{Shortest paths on a polyhedron}, in
  \emph{Proceedings of the Sixth Annual Symposium on Computational Geometry,
  Berkeley, CA, USA, June 6-8, 1990}, ed. R.~Seidel ({ACM}, 1990), pp.
  360--369.

\bibitem{ICH}
S.~Xin and G.~Wang, \enquote{Improving {Chen} and {Han's} algorithm on the
  discrete geodesic problem}, \emph{{ACM} Trans. Graph.} \textbf{28} (2009)
  104:1.

\bibitem{SVG}
X.~Ying, X.~Wang and Y.~He, \enquote{Saddle vertex graph {(SVG):} a novel
  solution to the discrete geodesic problem}, \emph{{ACM} Trans. Graph.}
  \textbf{32} (2013) 170:1.

\bibitem{Straightest}
K.~Polthier and M.~Schmies, \enquote{Straightest geodesics on polyhedral
  surfaces}, in \emph{International Conference on Computer Graphics and
  Interactive Techniques, {SIGGRAPH} 2006, Boston, Massachusetts, USA, July 30
  - August 3, 2006, Courses}, eds. J.~W. Finnegan and D.~Shreiner ({ACM},
  2006), pp. 30--38.

\bibitem{crane2018discrete}
K.~Crane, \enquote{Discrete differential geometry: An applied introduction},
  \emph{Notices of the AMS, Communication}  (2018) 1153.

\bibitem{bose2011survey}
P.~Bose, A.~Maheshwari, C.~Shu and S.~Wuhrer, \enquote{A survey of geodesic
  paths on {3D} surfaces}, \emph{Comput. Geom.} \textbf{44} (2011) 486.

\bibitem{crane2020survey}
K.~Crane, M.~Livesu, E.~Puppo and Y.~Qin, \enquote{A survey of algorithms for
  geodesic paths and distances}, \emph{CoRR} \textbf{abs/2007.10430}.

\bibitem{Surazhsky}
V.~Surazhsky, T.~Surazhsky, D.~Kirsanov, S.~J. Gortler and H.~Hoppe,
  \enquote{Fast exact and approximate geodesics on meshes}, \emph{{ACM} Trans.
  Graph.} \textbf{24} (2005) 553.

\bibitem{IVP}
P.~Cheng, C.~Miao, Y.~Liu, C.~Tu and Y.~He, \enquote{Solving the initial value
  problem of discrete geodesics}, \emph{Comput. Aided Des.} \textbf{70} (2016)
  144.

\bibitem{CyberTape}
C.~C.~L. Wang, \enquote{Cybertape: an interactive measurement tool on
  polyhedral surface}, \emph{Comput. Graph.} \textbf{28} (2004) 731.

\bibitem{martinez2005}
D.~M. Morera, L.~Velho and P.~C.~P. Carvalho, \enquote{Computing geodesics on
  triangular meshes}, \emph{Comput. Graph.} \textbf{29} (2005) 667.

\bibitem{xin2007}
S.~Xin and G.~Wang, \enquote{Efficiently determining a locally exact shortest
  path on polyhedral surfaces}, \emph{Comput. Aided Des.} \textbf{39} (2007)
  1081.

\bibitem{Flip}
N.~Sharp and K.~Crane, \enquote{You can find geodesic paths in triangle meshes
  by just flipping edges}, \emph{{ACM} Trans. Graph.} \textbf{39} (2020) 249:1.

\bibitem{CGAL}
{The CGAL Project}, \emph{{CGAL} User and Reference Manual} ({CGAL Editorial
  Board}, 2023), {5.5.2} edition.

\bibitem{Loop}
C.~Loop, \enquote{Smooth subdivision surfaces based on triangles, master's
  thesis}, \emph{University of Utah, Department of Mathematics} .

\end{thebibliography}

%%%%%%%%%%%%%%%%%%%%%%%%%%%%%%%%%%%%%%%%%

\end{document}